\documentclass[twoside,leqno,twocolumn]{article}

\usepackage{colortbl}

\usepackage{ltexpprt}
\usepackage{t1enc}
\usepackage[utf8]{inputenc}

\usepackage{amsfonts}

\newcommand{\abs}[1]{\left| #1\right|}

\newcommand{\set}[1]{\left\{ #1\right\}}

\newcommand{\sodass}{\,:\,}
\newcommand{\setGilt}[2]{\left\{ #1\sodass #2\right\}}

\newcommand{\natnull}{\mathbb{N}_{0}}

\newcommand{\realrange}[2]{\left[#1, #2\right]}

\newcommand{\unitrange}[2]{\realrange{0}{1}}

\newcommand{\Oh}[1]{\mathcal{O}\!\left( #1\right)}

\newcommand{\llabel}[1]{\label{\labelprefix:#1}}
\newcommand{\labelprefix}{} %

\newcommand{\discussionsize}{\small}

\newenvironment{code}{\noindent%
\begin{tabbing}%
\hspace{2em}\=\hspace{2em}\=\hspace{2em}\=\hspace{2em}\=\hspace{2em}\=%
\hspace{2em}\=\hspace{2em}\=\hspace{2em}\=\hspace{2em}\=\hspace{2em}\=%
\kill}{\end{tabbing}}

\newcommand{\labelcommand}{}
\newcommand{\captiontext}{}
\newsavebox{\codeparam}
\newcounter{lineNumber}
\newenvironment{disscodepos}[3]{%
\renewcommand{\labelcommand}{#2}%
\renewcommand{\captiontext}{#3}%
\sbox{\codeparam}{\parbox{\textwidth}{#3}}%
\begin{figure}[#1]\begin{center}\begin{code}\setcounter{lineNumber}{1}}{%
\end{code}\end{center}\caption{\llabel{\labelcommand}\captiontext}\end{figure}}

{\end{disscodepos}}

\newdimen\endofsize\endofsize=0.5em
\def\endofbeweis{~\quad\hglue\hsize minus\hsize
                 \hbox{\vrule height \endofsize width
\endofsize}\par}

\usepackage{epsfig}
\usepackage{url}
\usepackage{amsmath}
\usepackage{amssymb}
\usepackage{algorithm}
\usepackage[noend]{algpseudocode}
\usepackage{multicol}
\usepackage{pdflscape}
\usepackage{numprint}

\nprounddigits{2}
\npdecimalsign{.} %
\usepackage{makeidx}
\usepackage[usenames,dvipsnames,table,xcdraw]{xcolor}
\definecolor{lightergray}{rgb}{0.86, 0.86, 0.86}

\usepackage{hyperref}
\usepackage{xspace}
\usepackage{wrapfig}
\usepackage{graphics}
\usepackage{todonotes}
\usepackage{booktabs}
\usepackage{microtype}%
\definecolor{infocolor}{rgb}{0.6,0.6,0.6}

\newif\ifDoubleBlind
\DoubleBlindfalse

\usepackage{tikz}
\usetikzlibrary{shapes.arrows,chains}
 
\usetikzlibrary{decorations.pathreplacing}

\usepackage{framed,enumitem}

\usepackage[numbers,square,sort&compress]{natbib}

\newcommand{\Id}[1]{\texttt{\detokenize{#1}}}

\newcommand{\ori}[1]{\overline{#1}}
\DeclareMathOperator{\arboricity}{\alpha}

\newcommand{\ie}{i.\,e.,\xspace}
\newcommand{\eg}{e.\,g.,\xspace}
\newcommand{\etal}{et~al.~}

\usepackage{color}

\newcommand{\strash}[1]{{\color{orange}[DS: #1]}}

\newcommand{\lamm}[1]{{\color{blue}[SL: #1]}}
\newcommand{\sanders}[1]{{\color{blue}[PS: #1]}}
\newcommand{\werneck}[1]{{\color{blue}[RW: #1]}}
\renewcommand{\strash}[1]{}
\renewcommand{\lamm}[1]{}
\renewcommand{\sanders}[1]{}
\renewcommand{\werneck}[1]{}

\def\comment#1{}

\def\withcomments{
  \newcounter{mycommentcounter}
   \def\comment##1{\refstepcounter{mycommentcounter}%
    \ifhmode%
     \unskip%
     {\dimen1=\baselineskip \divide\dimen1 by 2 %
       \raise\dimen1\llap{\tiny\bfseries \textcolor{red}{-\themycommentcounter-}}}\fi%
     \marginpar[{\renewcommand{\baselinestretch}{0.8}%
       \hspace*{3em}\begin{minipage}{5em}\footnotesize [\themycommentcounter]: \raggedright ##1\end{minipage}}]{\renewcommand{\baselinestretch}{0.8}%
       \begin{minipage}{5em}\footnotesize [\themycommentcounter]: \raggedright ##1\end{minipage}}}
  }

\definecolor{darkgreen}{RGB}{0,200,100}
\definecolor{orange}{RGB}{255,80,0}
\definecolor{lightseagreen}{rgb}{0.13, 0.7, 0.67}

\newcommand{\Xcomment}[1]{}

\newcommand{\mytitle}{Engineering Fully Dynamic $\Delta$-Orientation Algorithms }

\begin{document}

\newcommand\relatedversion{}
\renewcommand\relatedversion{\thanks{The full version of the paper can be accessed at \protect\url{https://arxiv.org/abs/1902.09310}}} %

\title{\Large \mytitle}
\ifDoubleBlind
\else
\author{Jannick Borowitz, Ernestine Großmann, Christian Schulz} \fi

\date{}

\maketitle


\date{}

\maketitle
\begin{abstract}
A (fully) dynamic graph algorithm is a data structure that supports edge insertions, edge deletions, and answers certain queries that are specific to the problem under consideration.  There has been a lot of research on dynamic algorithms for graph problems that are solvable in polynomial time by a static algorithm.  However, while there is a large body of theoretical work on efficient dynamic graph algorithms, a lot of these algorithms were never implemented and   empirically evaluated.

In this work, we consider the fully dynamic edge orientation problem, also called fully dynamic $\Delta$-orientation problem, which is to maintain an orientation of the edges of an undirected graph such that the out-degree is low.  If edges are inserted or deleted, one may have to flip the orientation of some edges in order to avoid vertices having a large out-degree.  While there has been theoretical work on dynamic versions of this problem, currently there is no experimental evaluation available.  In this work, we close this gap and engineer a range of new dynamic edge orientation algorithms as well as algorithms from the current literature. Moreover, we evaluate these algorithms on real-world dynamic graphs.
The best algorithm considered in this paper in terms of quality, based on a simple breadth-first search, computes the optimum result on more than 90\%  of the instances and is on average only 2.4\% worse than the optimum solution.

\end{abstract}

\section{Introduction}
\label{sec:introduction}

Complex graphs are useful in a wide range of applications from technological networks to biological systems like the human brain.
These graphs can contain billions of vertices and edges. Often, analyzing these networks aids us in gaining new insights.
In practice, the underlying graphs often change over time, \ie vertices
or edges are inserted or deleted as time is passing.
In a social network, for example, users sign up or leave, and relations between
them may be created or removed over time, or in road networks new roads are built.
Terminology-wise, a problem is said to be \emph{fully dynamic} if the update
operations include both insertions \emph{and} deletions of edges.

A (fully) dynamic graph algorithm is a data structure that supports edge insertions, edge deletions, and answers certain queries that are specific to the problem under consideration.
The most studied dynamic problems are graph problems such as connectivity, reachability, shortest paths, or matching (see~\cite{DBLP:journals/corr/abs-2102-11169}).  
However, while there is a large body of theoretical work on efficient dynamic graph algorithms, most of these algorithms were never implemented and   empirically evaluated.
For some classical dynamic algorithms, experimental studies have been performed, such as early works on (all pairs) shortest paths \cite{DBLP:journals/jea/FrigioniINP98,DBLP:journals/talg/DemetrescuI06}, reachability~\cite{DBLP:conf/alenex/HanauerH020} or transitive closure~\cite{DBLP:journals/jea/KrommidasZ08,DBLP:conf/wea/HanauerH020} and later contributions for fully dynamic (graph, $k$-center) clustering~\cite{DBLP:conf/wads/DollHW11,DBLP:conf/alenex/GoranciHLSS21}, fully dynamic approximation of betweenness centrality~\cite{DBLP:conf/esa/BergaminiM15}, and fully dynamic minimum cuts~\cite{DBLP:conf/alenex/HenzingerN022}. However, while the engineering aspect in the area gathers some steam, still for a wide range of  other fundamental dynamic graph problems, theoretical algorithmic ideas have received very little attention from the practical perspective. In particular, very little work has been devoted to engineering such algorithms and providing efficient implementations in practice.

An important \emph{building block} of fully dynamic algorithms is to store sparse graphs with \emph{low} memory requirements while still being able to answer adjacency queries fast. The latter means that given two vertices $u$ and $v$ a function should return true if $\{u,v\} \in E$ and false otherwise -- in the best case this should take constant time. Traditional methods to store dynamic graphs use adjacency matrices, which need $O(n^2)$, but can answer such queries in $O(1)$ time, or storing adjacency arrays which require $O(n+m)$ space, but checking adjacency requires to search the complete neighborhood of a vertex which can potentially be large. 

For static graphs, Kannan \etal\cite{DBLP:journals/siamdm/KannanNR92} propose a method to efficiently store an undirected graph and support adjacency queries in time $O(\alpha)$ where $\alpha$ is the arboricity of a graph. 
Here, the \emph{arboricity} $\alpha(G)$ of a graph is defined as the smallest number $t$ such that $G$ can be partitioned into $t$ forests. 
The main idea of the algorithm is simple: each edge is stored in the list of only one of its two endpoints. Queries can then be performed by searching the adjacency lists of $u$ (for $v$) \emph{and} $v$ (for $u$).
By this trick both nodes can potentially have short adjacency lists, even if the vertices have a very high degree.
Storing an edge at only one of its endpoints is equivalent to assigning the edge an orientation (the edge is then stored at its start vertex).
More formal, an \emph{orientation} of a graph $G=(V,E)$ is a \emph{directed graph} $\ori{G}=(V,E')$ such that for every $\set{u,v}\in E$ either $(u,v)$ or $(v,u)$ is in $E'$.
To provide fast adjacency queries (constant-time), the out-degree of a vertex in $\ori{G}$ should be bounded by some constant $\Delta$ -- these orientations are called $\Delta$-orientations. 

In the static case, an $\alpha$-orientation always exists.
As stated by Nash-Williams~\cite{DBLP:journals/gc/ChenMWZZ94, 10.1112/jlms/s1-39.1.12, 10.1112/jlms/s1-36.1.445}, a graph $G$ has  arboricity~$\alpha$ if and only if $E$ can be partitioned into $E_1, \ldots ,E_{\alpha}$ such that $(V, E_i)$ is a forest. 
Computing an $\alpha$-orientation then works by choosing an arbitrary root for each forest and orienting the edges of the tree towards the root. Since each node is adjacent to at most $\alpha$ forest, the out-degree of the orientation is bounded by $\alpha$. Finding such tree decompositions is non-trivial, but can be done in polynomial time (see \cite{DBLP:journals/networks/PicardQ82,DBLP:journals/algorithmica/GabowW92}). In practice, one is thus interested in algorithms computing a $\Delta$-orientation with~$\Delta$~being small.

In the fully dynamic case, which is the focus of this work, maintaining $\Delta$-orientations and thus being able to support fast adjacency queries  is a very important building block of many dynamic graph algorithms \cite{DBLP:conf/icalp/KopelowitzKPS14} (see the full version of \cite{DBLP:conf/icalp/KopelowitzKPS14} for an even larger list).
For example, Neiman and Solomon \cite{DBLP:journals/talg/NeimanS16} have shown how to maintain a maximal matching in $O(\frac{\log n}{\log \log n})$ amortized time that uses a dynamic edge orientation algorithm, it is used in dynamic matrix vector multiplication~\cite{DBLP:conf/icalp/KopelowitzKPS14}, or Kowalki and Kurowski~\cite{DBLP:journals/talg/KowalikK06} use dynamic edge orientations to answer shortest-path queries of length at most $k$ in constant time in planar graphs. Other important applications of fully dynamic edge orientations include fully dynamic coloring~\cite{DBLP:conf/icalp/ChristiansenR22} and maintaining subgraph counts~\cite{DBLP:conf/sand/HanauerHH22}.
However, while there has been a wide-range of theoretical work on the edge orientation problem, none of the algorithms have been experimentally evaluated. 

\textbf{Contributions:} In this work, we close this gap and engineer a range of dynamic edge orientation algorithms from the current literature as well as new algorithms. Moreover, we evaluate and compare these algorithms on real-world dynamic graphs as well as dynamic graphs that have been obtained from real-world static graphs. We also develop an integer linear program to solve the problem to optimality on static graphs which enables use to solve a wide-range of static instances and to compare the dynamic algorithms to the optimum solution when all updates have been processed. Lastly, we give advice to practitioners.

\section{Preliminaries}
\label{sec:preliminaries}

\subsection{Basic Concepts.}
\label{subsec:basic_concepts}

Let $G=(V=\{0,\ldots, n-1\},E)$ be an \emph{undirected graph}. Let $\Gamma(v) = \setGilt{u}{\set{v,u}\in E}$ denote the neighbors of a vertex $v$ and $\deg(v)=\abs{\Gamma(v)}$ the \emph{degree of $v$}.
Further, let $\Delta(G)$ denote the maximum degree of $G$.
A graph-sequence $\mathcal{G} = (G_0, \ldots, G_t)$ for some $t\in\natnull$ is an \emph{edit-sequence of graphs} if there exists
for all $i>0$ an edge $e\in E(G_{i})$ such that it is either inserted, \ie $G_{i} = G_{i-1} + e$, or deleted, \ie $G_{i} = G_{i-1} - e$, in update~$i$.
The \emph{arboricity} $\arboricity(G)$ of a graph is defined as the smallest number $t$ such that $G$ can be partitioned into $t$ forests.
A graph-sequence $\mathcal{G}$ has \emph{bounded arboricity} $\alpha > 0$ if $\arboricity(G)\leq \alpha$ for all $G\in\mathcal{G}$.

An \emph{orientation} of a graph $G=(V,E)$ is a \emph{directed graph} $\ori{G}=(V,E')$ such that for every $\set{u,v}\in E$ either $(u,v)$ or $(v,u)$ is in $E'$. %
The out-degree of a node $u$ in a directed graph is defined as the number of edges starting in $u$, \ie $(u,v)\in E'$.
By $\Delta$ we refer to the maximum out-degree  of $v$ in $\ori{G}$.
$\ori{G}$ is a \emph{$c$-orientation} of $G$ if $\Delta\leq c$.
The graph-sequence $\ori{\mathcal{G}} = (\ori{G}_0, \ldots , \ori{G}_t)$ is a \emph{sequence of orientations} of $\mathcal{G}$ if every $\ori{G}_i$ is an orientation of $G_i$. 
In the same way, $\ori{\mathcal{G}}$ is a \emph{sequence of $c$-orientations} if all $\ori{G}_i$ are $c$-orientations.
The goal of the fully dynamic edge orientation problem is to keep the maximum degree $\Delta$ as small as possible at each point in time. 

 Given some orientation $\ori{G}$, we say, we \emph{flip} an edge $(u,v)\in E(\ori{G})$ if we delete it and insert $(v,u)$.
Let $P=(V_P,E_P)$ be a \emph{directed path}. $P$ is said to be a \emph{$u$-$v$-path} if it starts in $u$ and ends in $v$. The \emph{length} of $P$ is the number of edges in $P$.
A path $P$ is flipped by flipping every edge.

\subsection{Related Work.}
\label{subsec:related_work}
There is a wide range on fully dynamic algorithms in literature in general. The most studied dynamic problems are graph problems such as connectivity, reachability, shortest paths, or matching. We refer the reader to the recent survey \cite{DBLP:journals/corr/abs-2102-11169} for more details. Moreover, data structures for the problem of representing sparse \emph{static} graphs while providing constant time adjacency queries has been considered in a wide range of literature, \eg~\cite{DBLP:journals/dam/ArikatiMZ97,DBLP:conf/icalp/ChuangGHKL98,DBLP:journals/siamdm/KannanNR92,DBLP:conf/focs/MunroR97,DBLP:conf/wg/TalamoV98,DBLP:journals/dam/Turan84}. These methods focus on different aspect such as linear construction time, planar graphs or parallelization of the proposed algorithms, but are limited to the static case. For example, Kannan \etal\cite{DBLP:journals/siamdm/KannanNR92} propose a data structure to store a graph with low arboricity efficiently and supporting~$O(\alpha)$ time adjacency queries.

\textbf{Fully Dynamic Edge Orientation Algorithms.}
We now give a high-level overview of dynamic results in the literature. We give more details about the individual algorithms in Section~\ref{s:main}.

Brodal and Fagerberg \cite{DBLP:conf/wads/BrodalF99} were the first to consider the problem in the dynamic case. The authors present a linear space data structure for maintaining graphs with bounded arboricity. The data structure requires a bound $c$ on the arboricity of the graph as input. It then supports adjacency queries in $O(c)$ time, edge insertions in amortized time $O(1)$ as well as edge deletions in amortized time $O(c+\log n)$. The authors note that if the arboricity of a dynamic graph remains bounded, then the forest partitions may change due to the update. To deal with this the authors introduce a re-orientation operation, also called flipping above, which can change the orientation of an edge in order to maintain a small out-degree.
Kowalik \cite{DBLP:journals/ipl/Kowalik07} also needs a bound $c$ on the arboricity. In particular, Kowalik shows that the algorithm of Brodal and Fagerberg can maintain an $O(\alpha \log n)$ orientation of an initially empty graph with arboricity bounded by $c$ in $O(1)$ amortized time for insertions and $O(1)$ worst-case time for deletions. 
Kopelowitz \etal \cite{DBLP:conf/icalp/KopelowitzKPS14} gave an algorithm that does not need a bound on the arboricity as input. Their algorithm maintains an $O(\log n)$-orientation in worst-case update time $O(\log n)$ for any constant arboricity~$\alpha$.
He~\etal~\cite{DBLP:conf/isaac/HeTZ14} show how to maintain an $O(\beta \alpha)$-orientation in $O(\frac{\log(n/(\beta \alpha))}{\beta})$ amortized insertion time and $O(\beta \alpha)$ amortized edge deletion time thereby presenting a trade-off between quality of the orientation (the maximum out-degree) and the running time of the operations. Berglin and Brodal \cite{DBLP:journals/algorithmica/BerglinB20} gave an algorithm that allows a \emph{worst-case} user-specific trade-off between out-degree and running time of the operations. Specifically, depending on the user-specified parameters, the algorithm can main an $O(\alpha+\log n)$ orientation in $O(\log n)$ worst-case time or an $O(\alpha \log^2 n)$-orientation in constant worst-case time.
Recently, Christiansen~\etal\cite{DBLP:journals/corr/abs-2209-14087} published a report which contains algorithms that  make choices based entirely on local information, which makes them automatically adaptive to the current arboricity of the~graph. One of their algorithm maintains a $O(\alpha)$-orientation with worst-case update time $O(\log^2n \log \alpha)$. The authors also provide an algorithm with worst-case update time $O(\log n \log \alpha)$ to maintain an $O(\alpha + \log n)$-orientation.

\section{Delta-Orientation Algorithms}
\label{s:main}
We now give details of the algorithms that we consider in our experimental study. We start this section by (optimal) algorithms for static graphs. We use these algorithms later to compare various fully dynamic algorithms against optimum results. Note although one can compute an $\alpha$-orientation in polynomial time, it is unclear if the minimum edge orientation problem is in \texttt{NP} or \texttt{P}. To see that an $\alpha$-orientation is not necessarily optimal consider a cycle of length three. For this graph there is a 1-orientation, but the arboricity of the graph is two. Hence, we start this section with an optimum algorithm for the problem. 

Afterwards, we describe fully dynamic algorithms for the delta orientation problem. To this end, we engineer algorithms that have been theoretically studied before, e.g. the algorithm by Berglin and Brodal~\cite{DBLP:journals/algorithmica/BerglinB20} and the algorithm by Brodal and Fagerberg \cite{DBLP:conf/wads/BrodalF99}, as well as novel algorithms for the problem.

\subsection{(Optimal) Algorithms for Static Graphs.}
\label{sec:ilp}
Finding an optimum edge orientation can be done using the following integer linear program.
For each undirected edge $e \in E$, we create two variables $e_{(u,v)}, e_{(v,u)}$.
The interpretation is that the variable $e_{(u,v)}$ is one if the edge is oriented from $u$ to $v$ and zero otherwise. We now give the integer programming formulation:
\begin{align*}
    &\mathrm{minimize} \quad       \phi & &\\
    &\mathrm{s.t.}   \sum_{v\in V\setminus\set{u}} e_{(u,v)}  \leq \phi \quad \forall u\in V \\
    &   \quad \  \  e_{(u,v)} + e_{(v,u)} =1  \quad \ \ \ \forall \{u,v\} \in E  \\
    &   \quad \ \ e_{(u,v)}, e_{(v,u)} \in  \{0,1\}  \quad \forall \{u,v\} \in E  
\end{align*}
Here, the first constraint measures the outdegree of each vertex and makes sure that the outdegree of each vertex is smaller than the objective function value $\phi$.
The second constraint ensures that an edge $\{u, v\}$ is oriented either from $u$ to $v$ or in the opposite direction. Hence, a solution the program is a feasible edge orientation. As $\phi$ is minimized, we minimize the maximum out-degree. 

\textbf{Linear Programming Relaxation.} We can relax the integer program to become a linear program. This is done by dropping the integer constraints on the variable and instead adding constraints that limit the range of $e_{(u,v)}$ to $[0,1]$.
After running a linear program solver, we construct a feasible orientation by rounding, i.e. we pick the directed edge $(u,v)$ if $\lceil e_{(u,v)}\rceil = 1$.
If $e_{(u,v)} = e_{(v,u)} = 0.5$, we pick a random orientation of this edge. We do not experimentally evaluate this algorithm in our work, as our focus is on fully dynamic algorithms.

\subsection{Fully Dynamic Algorithms.}
We now present the details of the fully dynamic algorithms under consideration. All of these algorithm take an edit-sequence (insertion, deletion) of graphs as a stream of edges.
Most algorithms maintain the orientation of edges by storing out-neighbors in adjacency arrays \textsc{Adj}$[u]$ for the corresponding vertices. The only exception is the $K$-Flips algorithm which uses FIFO-queues instead.

\subsubsection{Improving $u$-$y$-Path Search.}
\begin{algorithm}[b!]
    \caption{Improving $u$-$y$-Path Search Algorithm}\label{algo:imprpathsearch}
    \begin{algorithmic}
        \Procedure{insertion}{$u$,$v$}
            \State \textsc{Adj}[$u$] $:=$ \textsc{Adj}[$u$] $\cup \{v\}$
            \If{deg($u) < \Delta$ \textbf{or} $\Delta = 1$} \Return
            \EndIf

\State \(\triangleright\) find a path $P = (u,\ldots,y)$,  $deg(y) < deg(u)-1$
            \State $P\gets$ \textsc{BFS-SEARCH}($u$) 
            \If{$P\not = \emptyset$}
                \State flip all edges of $P$
            \EndIf
        \EndProcedure
        \Procedure{deletion}{$u$,$v$}
            \State \textsc{Adj}[$u$] $:=$ \textsc{Adj}[$u$] $\backslash \{v\}$
            \State \textsc{Adj}[$v$] $:=$ \textsc{Adj}[$v$] $\backslash \{u\}$
        \EndProcedure
        \Procedure{Adjacent}{$u$,$v$}
            \State \Return ($u \in $\textsc{Adj}[$v$] \textbf{or} $v \in$\textsc{Adj}[$u$])
        \EndProcedure

    \end{algorithmic}
\end{algorithm}

The first algorithm maintains the current objective function value (max out-degree $\Delta$) in a global variable and maintains vertices by their degree in a bucket priority queue.
When an edge~$(u,v)$ is inserted, the algorithm assigns the edge to the adjacency list of $u$. It then checks if the degree $u$ is smaller than maximum degree or if the current maximum degree is one. 
If this is the case, then the algorithm directly returns.
Otherwise, the algorithm starts a breadth first search from $u$ in the directed graph induced by the current orientation to find a vertex $y$ with outdegree smaller than~deg$(u)-1$. The breadth first search algorithm stops as soon as it has found such a vertex.
Assume the algorithm finds such a vertex. Let the corresponding path be $p = (u, \ldots, y)$. Note that all edges on this path are oriented from $u$ to $y$. The algorithm flips each edge on the path, thereby increasing the degree of $y$ by one and decreasing the out-degree of $u$ by one.  Note that the flipping operations can be done in $O(|P|)$, as one can store the positions of the target vertices of the edges in the respective adjacency arrays while doing the breadth first search.

If no such vertex $y$ is found, then the algorithm continues with processing the next update operation.
When deleting $(u,v)$ we remove it from the corresponding adjacency arrays of the vertices $u$ and $v$. 
We give pseudocode for the insertion and deletion operation in~Algorithm~\ref{algo:imprpathsearch}.

The breadth first search can be implemented such that the complexity is bounded by the nodes and edges touched by the algorithm.
In our implementation, we limit the depth of the breadth first search to $d$, leaving the algorithm with a worst-case insertion time of $O(\Delta^{d})$.
The deletion operation runs in $O(\Delta)$ time.
\subsubsection{Random Paths.}

\begin{algorithm}[t!]
    \caption{Random Path Algorithm}\label{algo:randomwalk}
    \begin{algorithmic}
        \Procedure{insertion}{$u$,$v$}
            \State \textsc{Adj}[$u$] $:=$ \textsc{Adj}[$u$] $\cup \{v\}$
            \If{deg($u) < \Delta$ \textbf{or} $\Delta = 1$} \Return
            \EndIf
            \State $P\gets$ \textsc{Random-Path}($u$) 
            \If{$P\not = \emptyset$}
                \State flip all edges of $P$
            \EndIf

        \EndProcedure
        \Procedure{deletion}{$u$,$v$}
            \State \textsc{Adj}[$u$] $:=$ \textsc{Adj}[$u$] $\backslash \{v\}$
            \State \textsc{Adj}[$v$] $:=$ \textsc{Adj}[$v$] $\backslash \{u\}$
        \EndProcedure
    \end{algorithmic}
\end{algorithm}

So far we successively searched for feasible targets $y$ via a breadth first search starting from~$u$ when an edge $(u,v)$ was inserted.
In Algorithm~\ref{algo:randomwalk}, we try to find a vertex with smaller degree using a random path/walk.
As before, the algorithm maintains the current objective function value (max out-degree $\Delta$) in a global variable and maintains vertices by their degree in a bucket priority queue.
When an edge $(u,v)$ is inserted, the algorithm assigns the edge to the adjacency list of $u$. It then checks if the degree $u$ is smaller than maximum degree or the current objective function value is one. 
If this is the case, then the algorithm directly returns.
The random path-based algorithm works as follows: initially all nodes are unmarked. We start the path from vertex $y:=u$ and mark $y$. The algorithm then picks a random out-edge $(v, \tilde y)$ such that $\tilde y$ is not marked, and then continues by setting~$y$ to $\tilde y$ and marks $y$. If the degree of the current vertex is smaller than  deg$(u)-1$ the algorithm stops and flips the edges on the path. In this case, we found a path from $u$ to a vertex that can take one additional out-edge without increasing the current maximum.  In order to limit the running time of the algorithm, we restrict the maximum steps to $d$ for $d>0$. The worst case running time for insertion is $O(d)$. 
In our implementation,  we repeat the path search (at most) $r$ times if the algorithm did not find a path and stop the path search as soon as we found one, yielding a worst-case update time of $O(rd)$.
Deletion works as in the breadth first case: 
 deleting $(u,v)$ removes it from the corresponding adjacency arrays of the vertices $u$ and $v$ and runs in $O(\Delta)$ time.

\subsubsection{Descending Degrees.} The core idea of the next algorithm is to always try to swap an edge $(u,v)$ with the neighbor that has the smallest degree.
As before, the algorithm maintains the current objective function value (max out-degree $\Delta$) in a global variable and maintains vertices by their degree in a bucket priority queue.
As before, when an edge $(u,v)$ is inserted, the algorithm assigns the edge to the adjacency list of $u$. It then checks if the degree $u$ is smaller than maximum degree. 
If this is the case, then the algorithm directly returns. Otherwise, it looks for the node $v$ with the smallest degree in $\textsc{Adj}$[$u$]. If the degree of $v$ is smaller than deg($u)-1$ then the algorithm swaps the edge and continues the process at $v$ recursively until we arrive at a node for which there is no such swap. If the initial swap at $u$ was successful, we repeat the overall process. The deletion of an edge works as before. Pseudocode can be found in Algorithm~\ref{algo:DecDegree}.

The worst-case running time of the insert routine is as follows: first lets look at the algorithm without the while loop. A single call of descending degrees starting at $u$ has running time $O(\Delta^2)$. Scanning the neighborhood of a node takes $O(\Delta)$ time, and assuming it was successful, also requires the degree of the next vertex to be at least one smaller. Hence, the overall path discovered by the algorithm can have length at most $O(\Delta)$. As the while loop is only continued if the degree of $u$ has been decreased by one, it is also bounded by $O(\Delta)$ iterations. Hence, the overall running time for an insert operation is $O(\Delta^3)$.

\subsubsection{Naive.} The naive algorithm always orients the inserted edge $(u,v)$ starting from the node with the smaller degree and ending in the node with the larger degree. No additional data structure to maintain the current maximum degree vertices is required. The running time of the algorithm is $O(1)$. Deletion works as before in $O(\Delta)$.
\begin{algorithm}[t]
    \caption{Descending Degrees Algorithm}\label{algo:DecDegree}
    \begin{algorithmic}

        \Procedure{insertion}{$u$, $v$}
            \State \textsc{Adj}[$u$] $:=$ \textsc{Adj}[$u$] $\cup\{v\}$
            \If{deg($u) < \Delta$ \textbf{or} $\Delta = 1$} \textbf{return} \EndIf
            \State \textbf{while} \textsc{DescendingDegrees}($u$) ;
        \EndProcedure
    \end{algorithmic}
    \begin{algorithmic}
        \Procedure{DescendingDegrees}{$u$}
            \State $v := $argmin$_{v \in \textsc{Adj}[u]} \text{deg}(v)$
            \State \textbf{if} $\text{deg}(v) < \text{deg}(u) - 1$ \textbf{then} 
            \State \quad \textsc{Adj}$[u] := \textsc{Adj}[u] \backslash \{v\}$
            \State \quad \textsc{Adj}$[v] := \textsc{Adj}[v] \cup \{u\}$
            \State \quad \textsc{DescendingDegrees}($v$);
            \State \quad \textbf{return} true;
            \State \textbf{return} false;

        \EndProcedure
    \end{algorithmic}
    \begin{algorithmic}
        \Procedure{deletion}{$u$,$v$}
            \State \textsc{Adj}[$u$] $:=$ \textsc{Adj}[$u$] $\backslash \{v\}$
            \State \textsc{Adj}[$v$] $:=$ \textsc{Adj}[$v$] $\backslash \{u\}$
        \EndProcedure
    \end{algorithmic}
\end{algorithm}

\subsubsection{\textsc{$K$-Flips}.}
\begin{algorithm}[t!]
    \caption{\textsc{$K$-Flips}  \cite{DBLP:journals/algorithmica/BerglinB20}}\label{algo:kflip}
    \begin{algorithmic}
        \Procedure{insertion}{$u$,$v$}
            \State push $(u,v)$ to $Q_u$ 
            \State \Call{k-flips}{}
        \EndProcedure
        \Procedure{deletion}{$u$,$v$}
            \State remove $(u,v)$ from $Q_u$
            \State \Call{k-flips}{}
        \EndProcedure
    \end{algorithmic}
    \begin{algorithmic}
        \Procedure{$K$-flips}{}
            \For{$i=1$ to $k$}
                \State $u :=$ a max out-degree vertex
                \State remove $(u,v)$ from $Q_u$
                \State push $(v,u)$ to $Q_v$
            \EndFor
        \EndProcedure
    \end{algorithmic}
\end{algorithm}

Berglin and Brodal \cite{DBLP:journals/algorithmica/BerglinB20} presented a simple greedy algorithm that maintains a bound 
on the maximum out-degree using a fixed number of flip operations.
The algorithm maintains for each vertex $v$ a standard FIFO queue $Q_v$, which holds all of its out-edges.
We give the pseudocode in Algorithm \ref{algo:kflip}.
The \textsc{$K$-flip} operation is always performed after inserting and deleting an edge $(u,v)$.
The \textsc{$K$-flip} operation performs $k$ flips of edges $(x,y)$ where $x$ has maximum out-degree in the current orientation.
We maintain the vertices of same out-degree in a bucket priority queue, \ie a degree indexed array and a pointer to the bucket corresponding to the maximum degree.

The $K$-Flips  algorithm allows a \emph{worst-case} user-specific trade-off between out-degree and running time of the operations via the parameter $k$. Specifically, depending on the user-specified parameters, the algorithm can main an $O(\alpha+\log n)$ orientation in $O(\log n)$ worst-case time or an $O(\alpha \log^2 n)$-orientation in constant worst-case time. More precisely, for $k=4$ the algorithm maintains an $\Oh{\alpha \log^2(n)}$-orientation  and for $k=\Oh{\log(n)}$ the algorithm maintains an $\Oh{\alpha + \log(n)}$-orientation.

\subsubsection{Algorithms For Known $\alpha$.}

\begin{algorithm}[t]
    \caption{Brodal and Fagerberg \cite{DBLP:conf/wads/BrodalF99}}\label{algo:brodal}
    \begin{algorithmic}

 \State \(\triangleright\) \textbf{Input:} $\tilde \alpha$ a bound on arboricity $\alpha$ 
        
        \Procedure{insertion}{$u$, $v$}
            \State \textsc{Adj}[$u$] $:=$ \textsc{Adj}[$u$] $\cup\{v\}$
            \If{$\deg(u) = \tilde \alpha + 1$}
                \State $S := \set{u}$
                \While{$\abs{S}\not = \emptyset$}
                    \State $w :=$ pop($S$)
                    \For{$x\in \textsc{Adj}[w]$}
                        \State $\textsc{Adj}[x] := \textsc{Adj}[x]\cup\set{w}$
                        \If{$\deg(x) = \tilde \alpha + 1$}
                            \State push($S, x$)
                        \EndIf
                    \EndFor
                    \State $\textsc{Adj}[w] := \emptyset$
                \EndWhile
            \EndIf
        \EndProcedure
    \end{algorithmic}
    \begin{algorithmic}
        \Procedure{deletion}{$u$,$v$}
            \State \textsc{Adj}[$u$] $:=$ \textsc{Adj}[$u$] $\backslash \{v\}$
            \State \textsc{Adj}[$v$] $:=$ \textsc{Adj}[$v$] $\backslash \{u\}$
        \EndProcedure
    \end{algorithmic}
    \begin{algorithmic}
        \Procedure{Rebuild}{($V, E$)}
            \State $\tilde \alpha := 2 \cdot \tilde \alpha$
            \State \textbf{for} $v \in V$ \textbf{do} \textsc{Adj}[$v$] $:= \emptyset$
            \State \textbf{for} $(u,v) \in E$ \textbf{do} \textsc{Insertion}($u,v$)
        \EndProcedure
    \end{algorithmic}

\end{algorithm}

Finally, we look at algorithms which need an upper bound $\tilde \alpha$ on the arboricity as input. This requirement implies that such an algorithm has knowledge about the whole edit-sequence. 
The algorithm by Brodal and Fagerberg \cite{DBLP:conf/wads/BrodalF99} is designed as such. 
We give pseudocode in Algorithm \ref{algo:brodal}.
After an insertion of an edge $(u,v)$ the algorithm proceeds as follow. If the out-degree of $u$ exceeds $\tilde\alpha$, the algorithm repeatedly picks a node $w$ with outdegree larger than $\tilde \alpha$ and the orientation of \emph{all} out-going edges is changed. Afterwards $w$ has out-degree zero. The process continues until all nodes have out-degree at most $\tilde \alpha$. The delete operation removes the corresponding directed edge. %
Brodal and Fagerberg \cite{DBLP:conf/wads/BrodalF99} show that the number of edge flips is finite if one chooses $\tilde \alpha \geq 2\max_{i=0}^t{\arboricity(G_i)}$.
Overall, the algorithm supports edge insertions in amortized time $O(1)$ as well as edge deletions in amortized time $O(\tilde \alpha+\log n)$. 

The authors also sketch an adaptive variant that does not need to know $\tilde \alpha$. In this case, the current graph is build again where $\tilde \alpha := 2 \tilde \alpha$ if a counter for edge-reorientations exceeds a bound depending on the current $\tilde\alpha$. In our implementation, we start with $\tilde\alpha = 1$ and always add $\tilde \alpha + 1$ on a bound for the number of reorientations (as described in \cite{DBLP:conf/wads/BrodalF99}) before rebuild has to be called.  The insert operation then takes $O(\log \tilde \alpha)$ amortized time. In our experiments, we also use other factors $\tilde \alpha := \beta \tilde\alpha$ with $\beta \in [1,2]$ in order to obtain better results at the expense of additional running time.

\ifDoubleBlind
\else
\subsection{Data Reduction / Pruning.} In our algorithms above, we used two pruning techniques. First, if the current update did not create a vertex with maximum degree in the network then we skipped the update (as we did not change the objective function with the newly inserted edge). On the other hand, we skipped the update if the maximum degree in the network has been equal to one. This is done, because there can not be any better solution than this as long as there is one edge in the network. For some specific applications of the problem, it should be possible to prune even more update steps. For example, for adjacency queries in dynamic algorithms, it may not be necessary to fully optimize/minimize the maximum outdegree, since the method to check adjacency between two vertices will most often have at least one cache miss. Thus one could potentially skip updates as long as the two adjacency arrays of the nodes considered in the update fit into the cache. 
\fi
\section{Experimental Evaluation}
\label{sec:experiments}
\label{s:exp}

\noindent \textbf{\textit{Methodology.}}
We performed our implementations using C++ and compiled them using gcc 8.3 with full optimization turned on (-O3 flag). 
All of our experiments were run on a single core of a  machine 
equipped with one Intel Xeon Silver 4216 16-Core CPU running at 3.20GHz with 22MB L3 Cache   96GB of main memory.  
The machine runs Ubuntu 20.04.1 LTS.

By default we perform ten repetitions per instance.
We measure the \emph{total time} taken to compute \emph{all edge insertions and deletions}
and generally use the \emph{geometric mean} when averaging over different instances
in order to give every instance a comparable influence on the final result. 
In order to compare different algorithms, we use \emph{performance profiles}~\cite{DBLP:journals/mp/DolanM02}.
These plots relate the objective function size / running time  of all algorithms to the corresponding objective function size / running time produced / consumed by each algorithm.
More precisely, the $y$-axis shows $\#\{\text{objective} \leq \tau * \text{best} \} / \# \text{instances}$, where objective corresponds to
the result of an algorithm on an instance and best refers to the best result of any algorithm shown within the plot.
When we look at running time, the $y$-axis shows $\#\{t \leq \tau * \text{fastest} \} / \# \text{instances}$, where $t$ corresponds to
the time of an algorithm on an instance and fastest refers to the time of the fastest algorithm on that instance.
The parameter $\tau\geq 1$ in this equation is plotted on the $x$-axis.
For each algorithm, this yields a non-decreasing, piecewise constant function.
Thus, if we are interested in the number of instances where an algorithm is the best/fastest, we only need to look at $\tau = 1$.

\noindent \textbf{\textit{Instances.}} We evaluate our algorithms on a number of large graphs. 
These graphs are collected from
      \cite{benchmarksfornetworksanalysis,DBLP:journals/corr/abs-2003-00736,UFsparsematrixcollection,snap,DBLP:conf/www/Kunegis13,konect:unlink,DBLP:journals/jpdc/FunkeLMPSSSL19,kappa}.
      Table~\ref{tab:graphstable} summarizes the main properties of the benchmark set.
      Our benchmark set includes a number of graphs from numeric simulations as well as complex networks.
These include static graphs as well as real dynamic graphs.
The graphs \Id{rhg*} are complex networks generated with KaGen~\cite{DBLP:journals/jpdc/FunkeLMPSSSL19} according to the \emph{random hyperbolic graph} model~\cite{Krioukov2010}. In this model nodes are represented as points in the hyperbolic
plane; nodes are connected by an edge if their hyperbolic distance is below a threshold.
Moreover, we use the two graph families \Id{rgg} and \Id{del} for
comparisons. 
\Id{rgg}$X$is a \emph{random geometric graph} with
$2^{X}$ nodes where nodes represent random points in the (Euclidean) unit square and edges
connect nodes whose Euclidean distance is below $0.55 \sqrt{ \ln n / n }$.
This threshold was chosen in order to ensure that the graph is almost certainly connected. 
\Id{del}$X$ is a Delaunay triangulation of $2^{X}$
random points in the unit square. 
These graphs have been generated using code from \cite{kappa}.

As our algorithms do only handle undirected graphs, we consider all input graphs to be undirected by ignoring edge directions and we remove self-loops and parallel edges.
We use the algorithms using insertions only (as the deletion routines are the same for all algorithms). For static graphs,  we start with an empty graph and insert all edges of the static graph in a random order. 
We also use real dynamic instances from Table~\ref{dyninstances} -- most of these instances, however, only feature insertions (with the exception being \texttt{dewiki} and \texttt{wiki-simple-en}) as there is currently a lack if publicly available instances that also feature deletions.

\subsection{Parameter Study.} \label{sec:parameterstudy} We now look at the parameters of all algorithms on all instances individually. All of the parameters of the algorithms are exhaustive, i.e.~larger (or smaller depending on the parameter) values of the respective parameter typically yields better quality (smaller maximum degree in the orientation) at the expense of higher running time per update. 

\paragraph*{Algorithms without parameters.} The naive and the descending degrees algorithm have no parameters. Thus we run them as they are in the Section \ref{exp:overall}.

\begin{figure*}[t!]
\centering
\vspace*{-1cm}
\includegraphics[width=7.25cm]{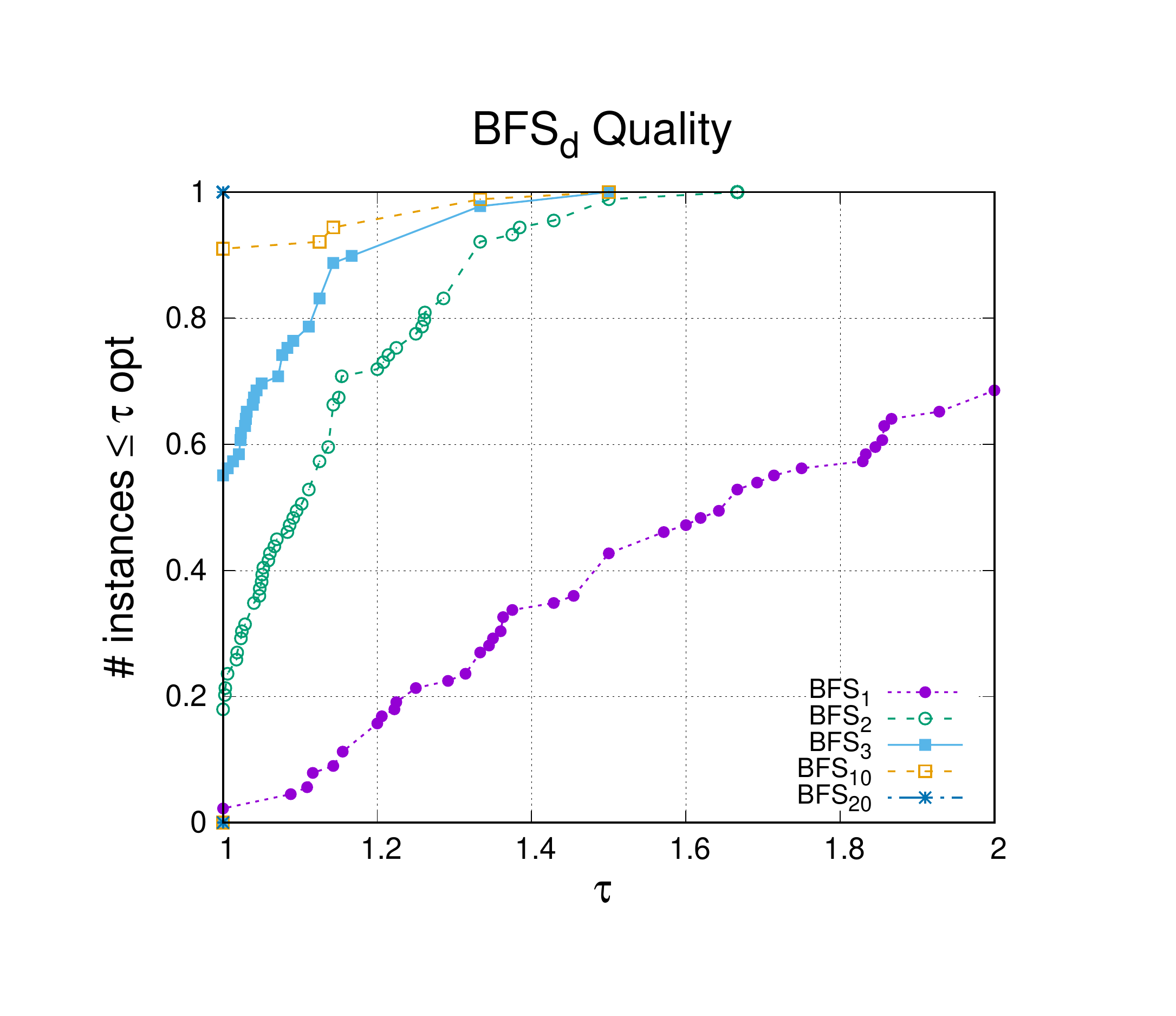}
\includegraphics[width=7.25cm]{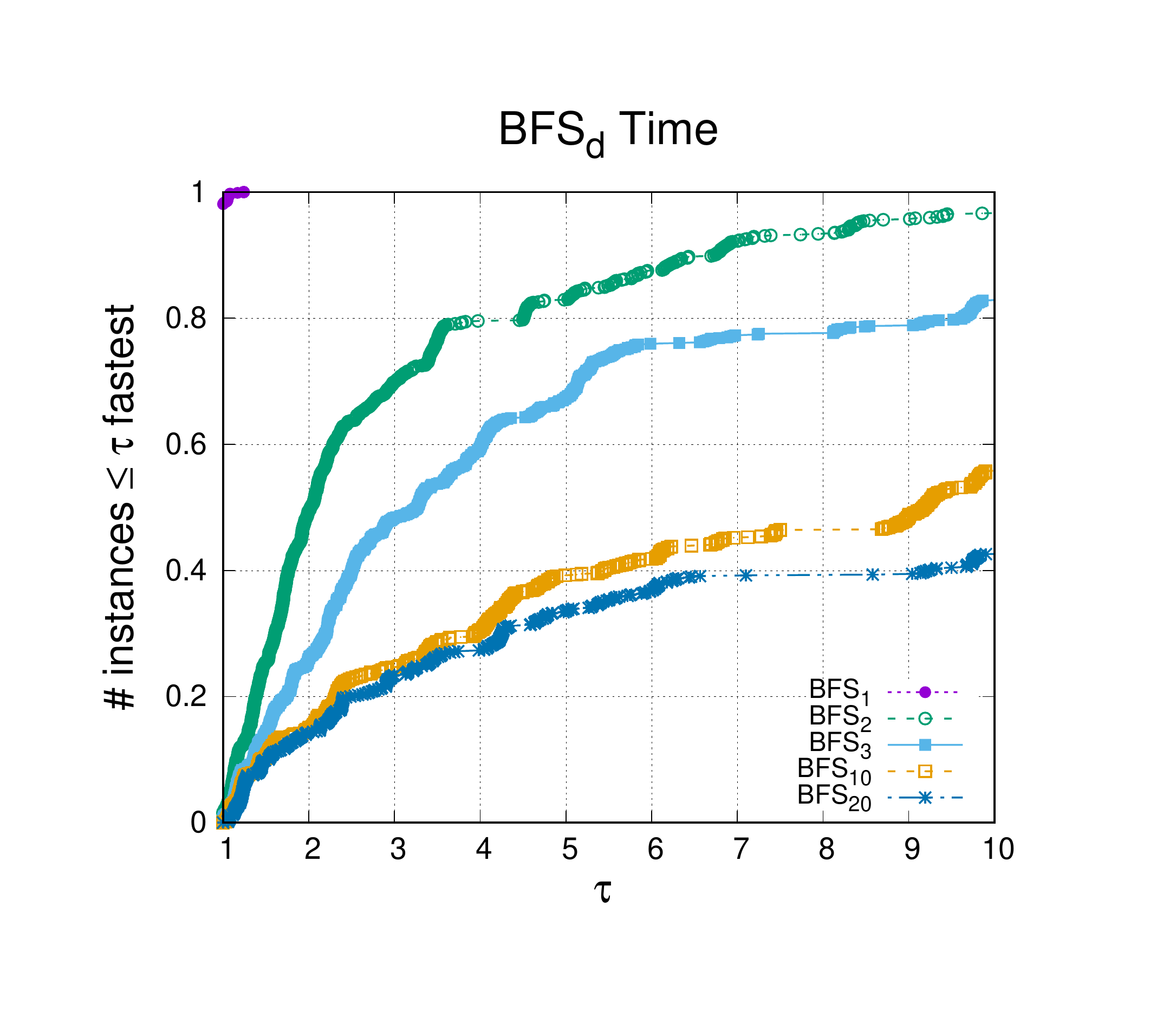}
\vspace*{-1cm}

\includegraphics[width=7.25cm]{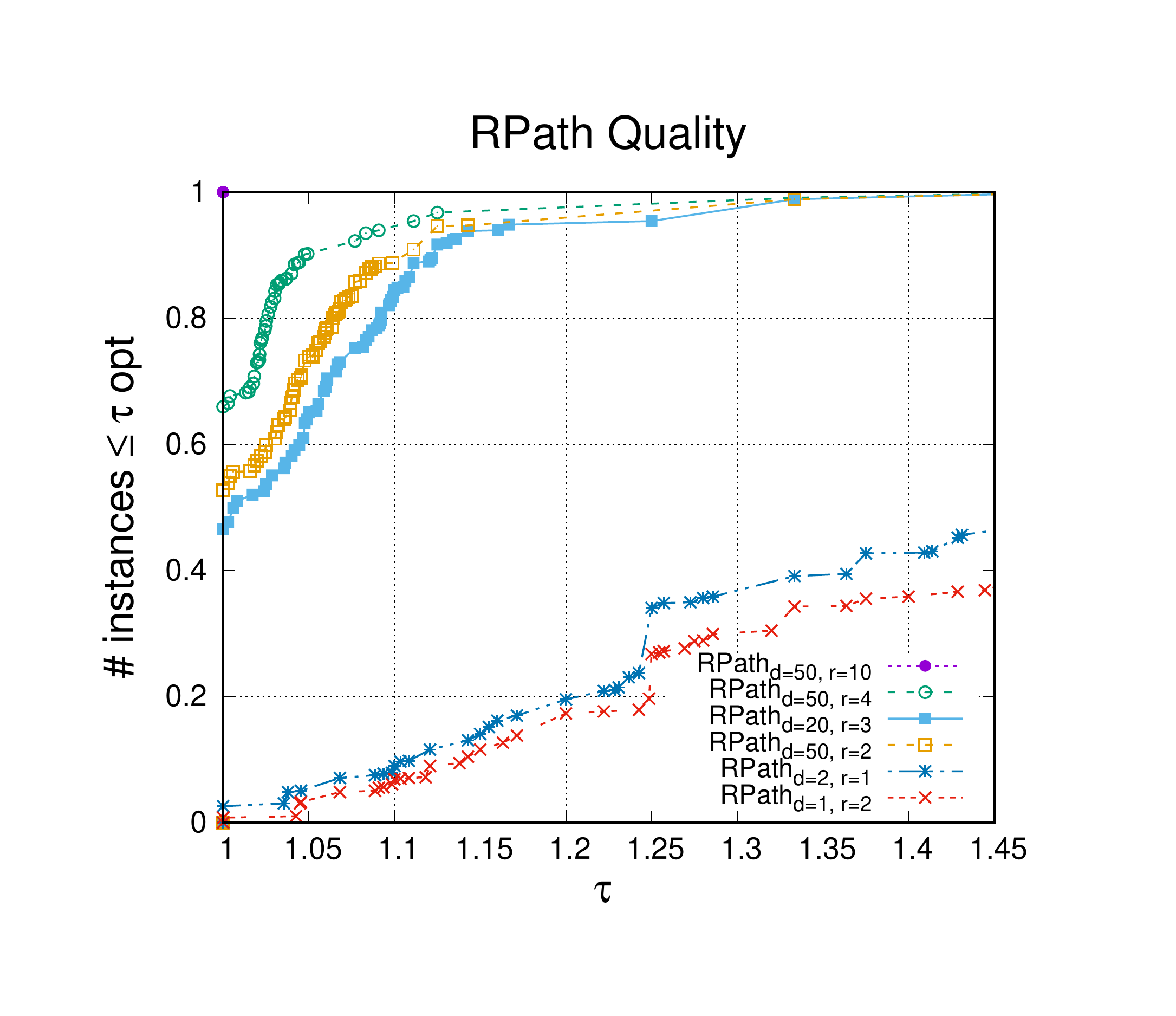}
\includegraphics[width=7.25cm]{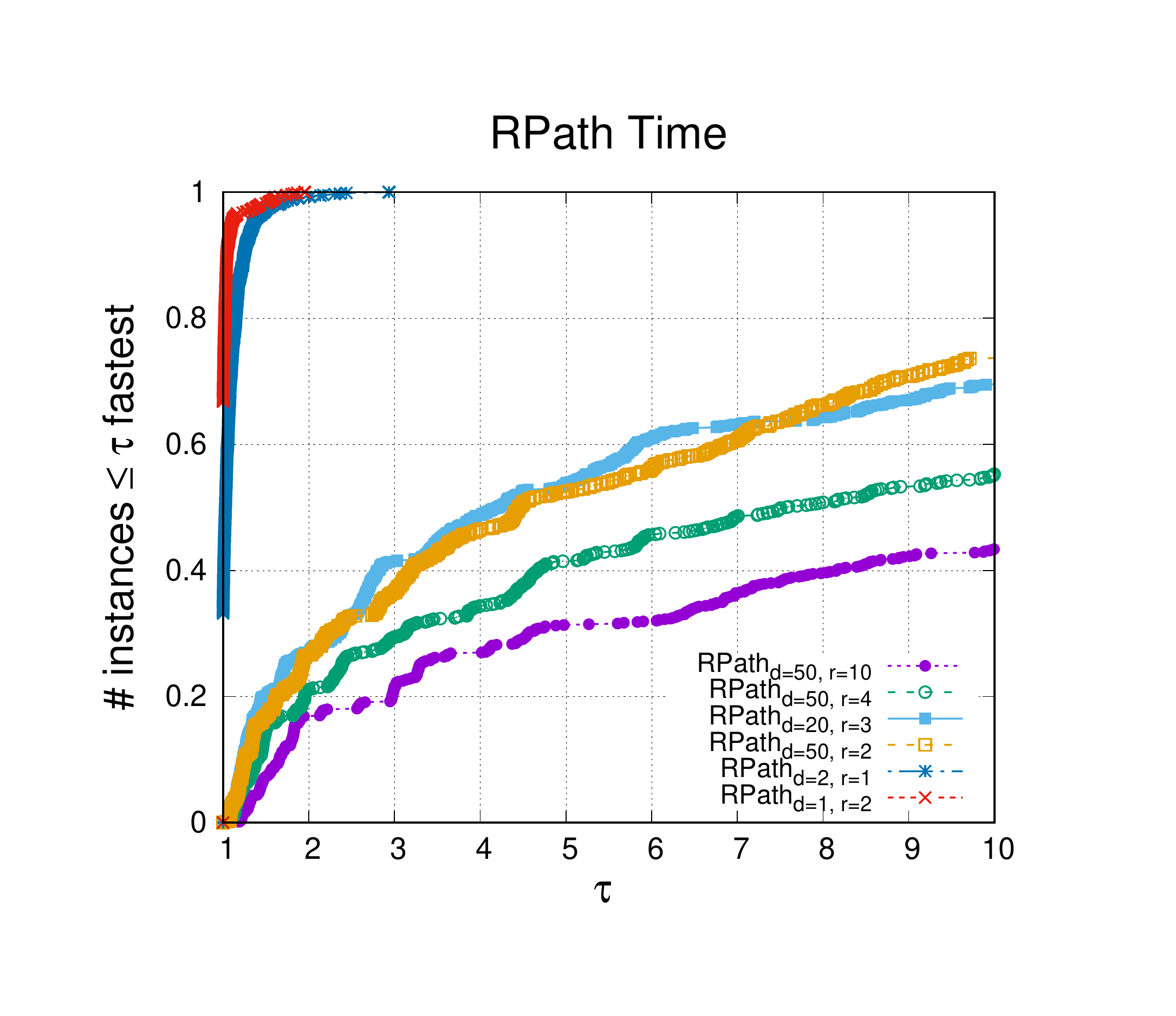} \\
\vspace*{-1cm}

\includegraphics[width=7.25cm]{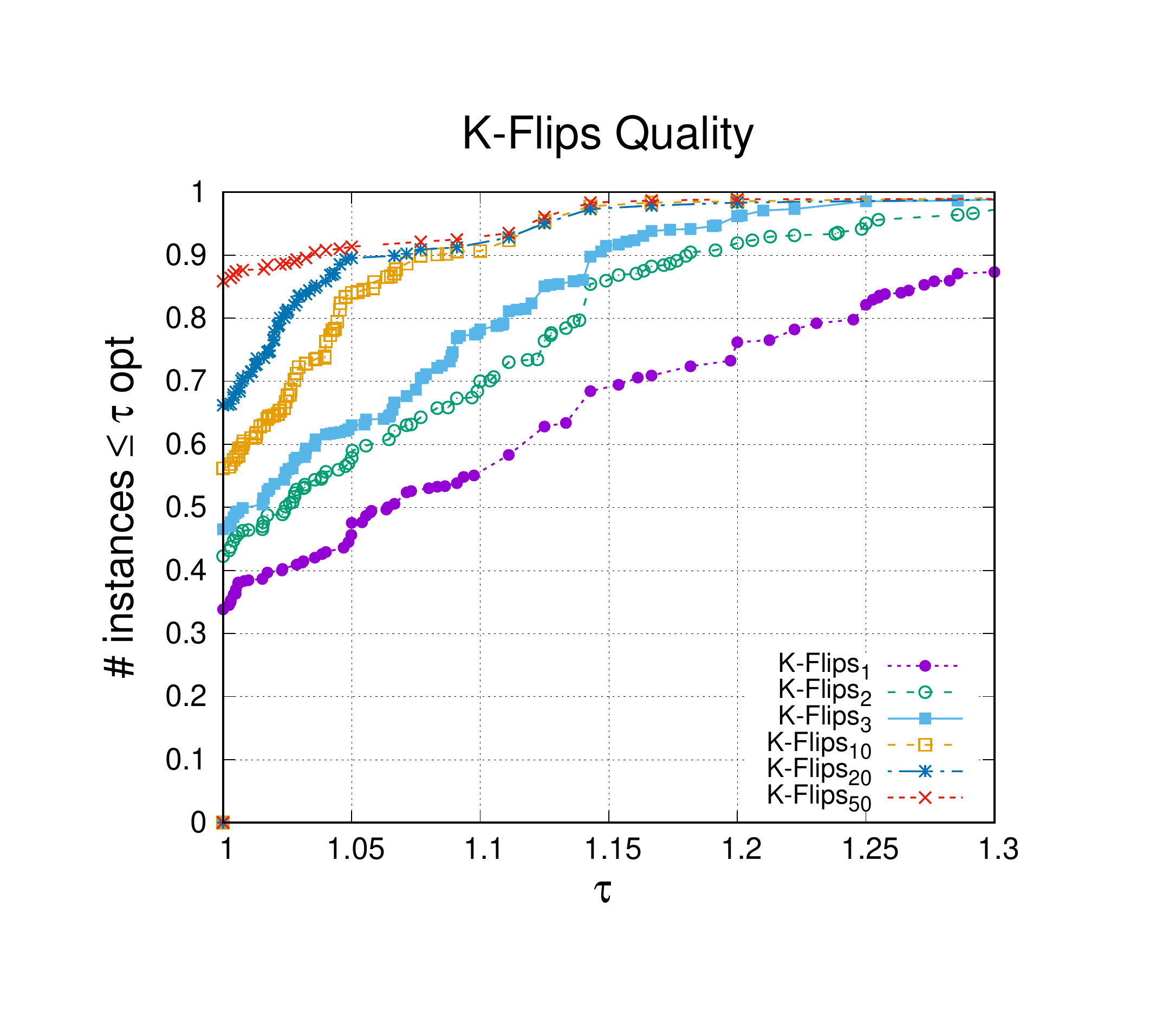}
\includegraphics[width=7.25cm]{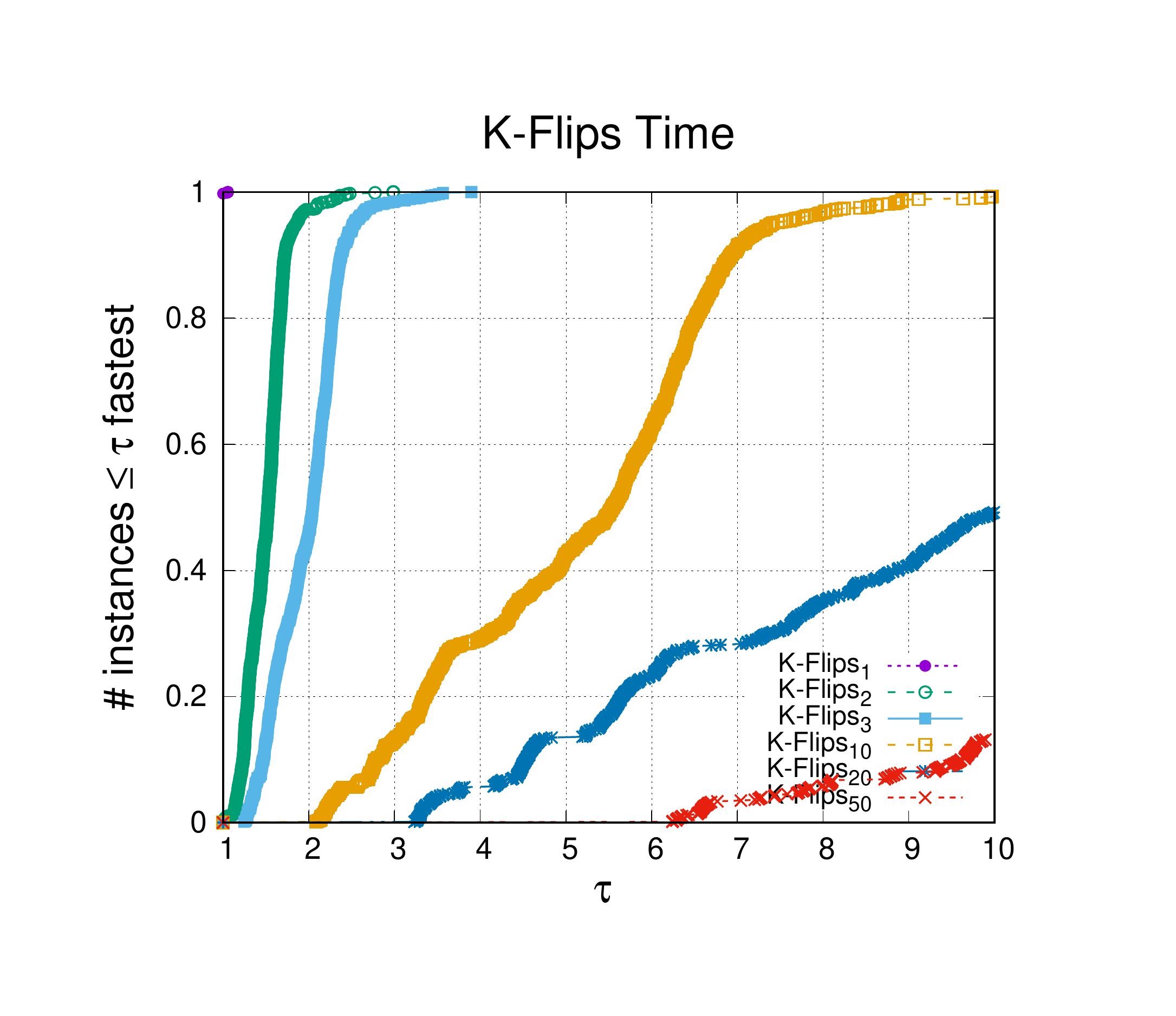} \\
\vspace*{-1cm}
\includegraphics[width=7.25cm]{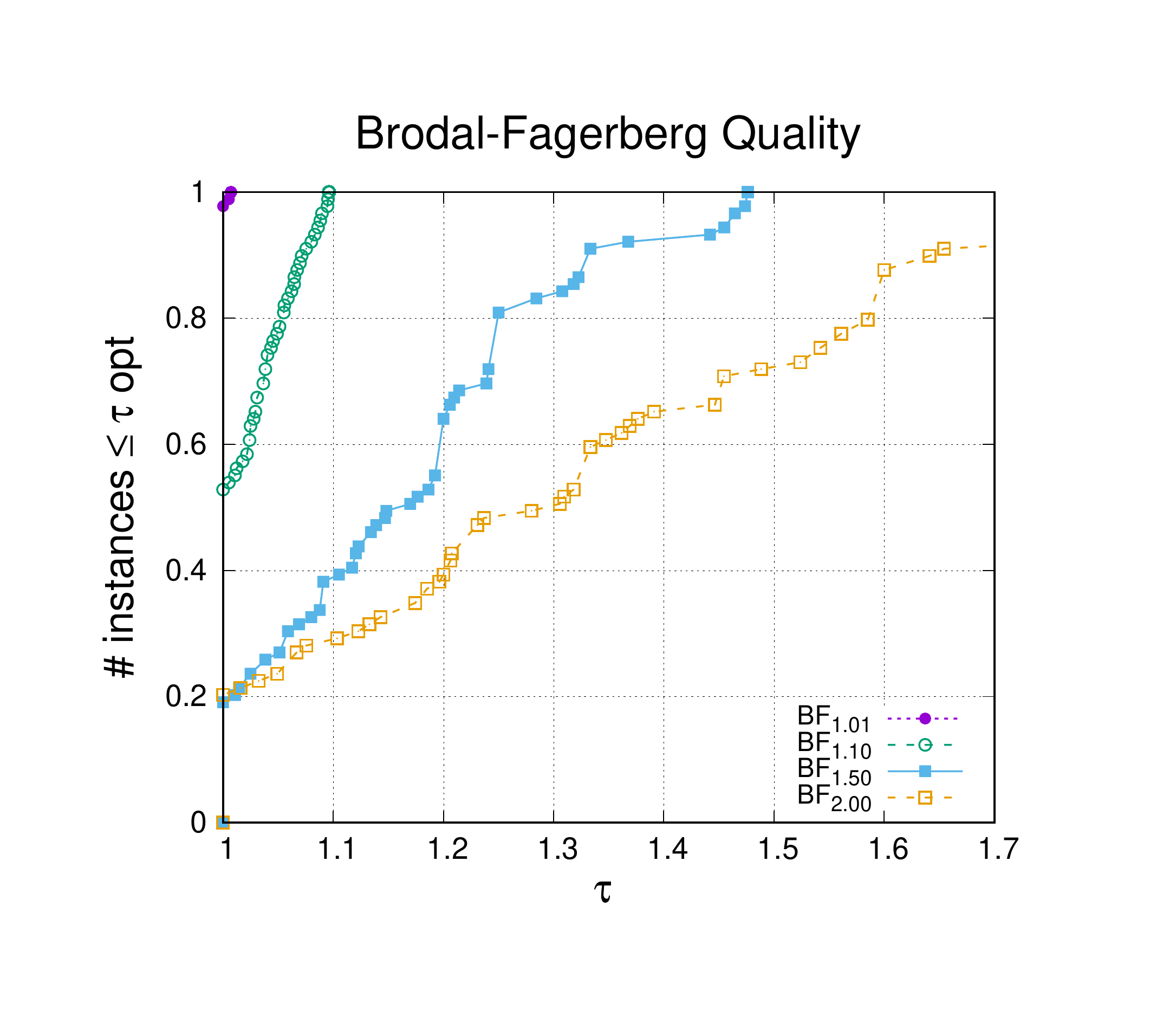}
\includegraphics[width=7.25cm]{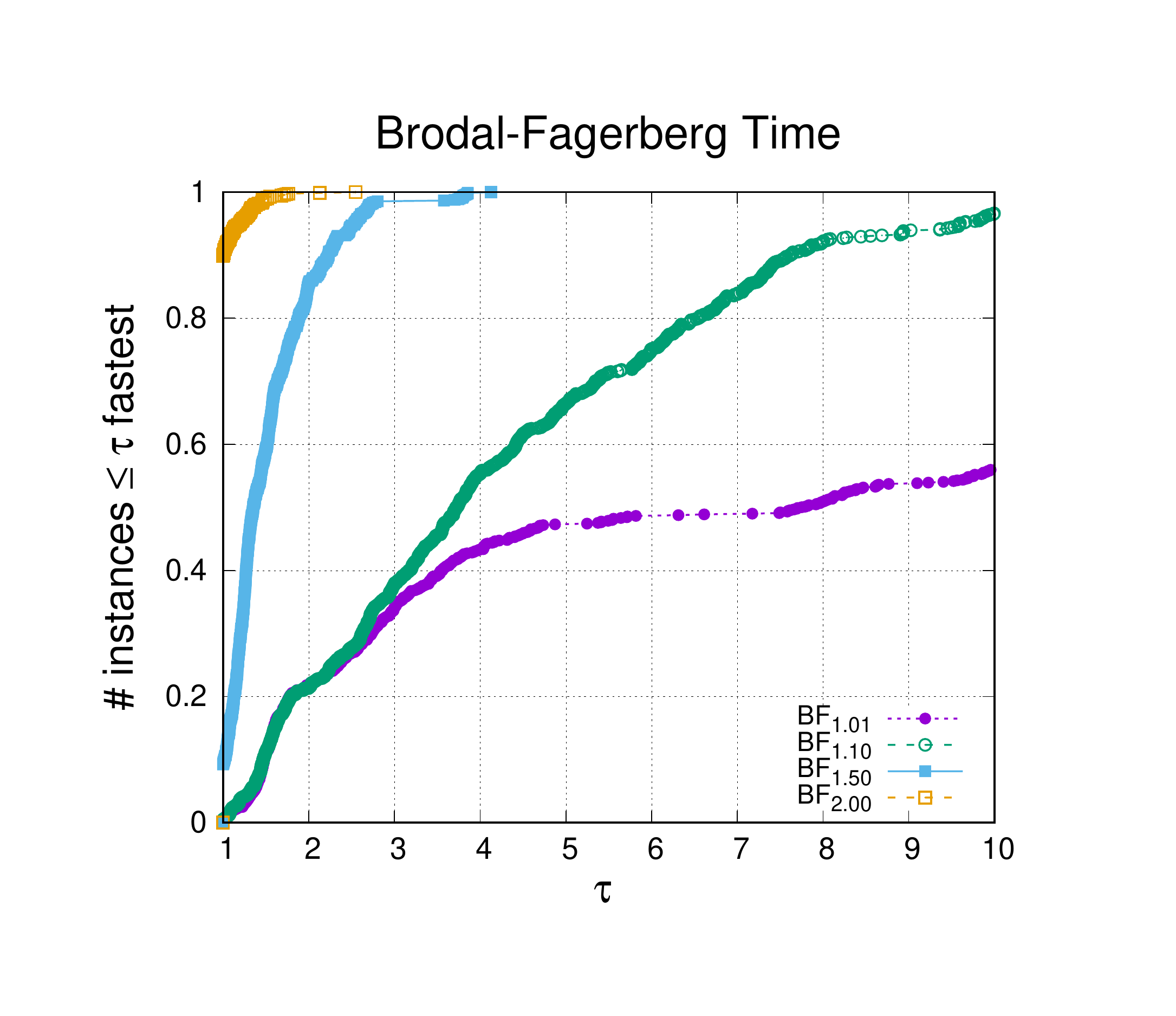}

\vspace*{-1cm}
\caption{Performance profile for solution quality (left) and running time (right) for algorithm configurations of (top to bottom) \texttt{BFS} (depth), \texttt{RPath} (depth, \#repetitions), \texttt{K-Flips} (\#flips) and \texttt{Brodal-Fagerberg} (scaling factor~$\beta$)~algorithms.}

\label{fig:expkflips}
\label{fig:perfbfs}
\label{fig:exprandompath}
\label{fig:expbf}
\end{figure*}
\begin{table*}[t!]
\centering
\caption{Average performance of all configurations considered sorted by running time for the overall update sequence on all instances. Pareto optimal configurations in terms of average values over all instances highlighted in \noindent\colorbox{lightergray} {\parbox{\widthof{gray}}{gray}}.}
\vspace*{0.25cm}

\begin{tabular}{lrr@{\hskip 43pt} lrr@{\hskip 23pt}}

\toprule
Algorithm & Avg $t$ [s] & Avg $\Delta$ & Algorithm &  Avg $t$ [s] & Avg $\Delta$\\
                  \midrule

\cellcolor{lightergray}\texttt{Naive}               & \numprint{0.0346441} & \numprint{25.0252} & \cellcolor{lightergray}\texttt{RPath}$_{d=50, r=2}$  & \numprint{0.395678} & \numprint{21.0562} \\
\texttt{RPath}$_{d=1, r=2}$                         & \numprint{0.0848061} & \numprint{34.0571} & \cellcolor{lightergray}\texttt{BFS}$_3$              & \numprint{0.481005} & \numprint{20.7274} \\
\texttt{RPath}$_{d=2, r=1}$                         & \numprint{0.0895722} & \numprint{33.1626} & \texttt{BF}$_{1.10}$          & \numprint{0.525975} & \numprint{27.3307} \\
\cellcolor{lightergray}\texttt{DescDegrees}   & \numprint{0.0909213} & \numprint{21.2435} & \texttt{$K$-Flips}$_{10}$     & \numprint{0.543161} & \numprint{21.2241} \\
\texttt{$K$-Flips}$_1$                              & \numprint{0.111624}  & \numprint{23.1999} & \cellcolor{lightergray}\texttt{RPath}$_{d=50, r=4}$  & \numprint{0.60334}  & \numprint{20.6976} \\
\texttt{BFS}$_1$                                    & \numprint{0.126289}  & \numprint{35.2697} & \texttt{$K$-Flips}$_{20}$     & \numprint{0.967601} & \numprint{21.1062} \\
\texttt{BF}$_{2.00}$                                & \numprint{0.149275}  & \numprint{34.5357} & \cellcolor{lightergray}\texttt{RPath}$_{d=50, r=10}$ & \numprint{1.06172}  & \numprint{20.2353} \\
\texttt{$K$-Flips}$_2$                              & \numprint{0.16643}   & \numprint{21.9994} & \texttt{BF}$_{1.01}$          & \numprint{1.17361}  & \numprint{26.7108} \\
\texttt{BF}$_{1.50}$                                & \numprint{0.213032}  & \numprint{30.945}  & \cellcolor{lightergray}\texttt{BFS}$_{10}$           & \numprint{1.32274}  & \numprint{19.9935} \\
\texttt{$K$-Flips}$_3$                              & \numprint{0.216885}  & \numprint{21.7003} & \texttt{$K$-Flips}$_{50}$     & \numprint{2.21084}  & \numprint{20.9596} \\
\texttt{BFS}$_2$                                    & \numprint{0.304544}  & \numprint{22.1265} & \cellcolor{lightergray}\texttt{BFS}$_{20}$           & \numprint{2.34556}  & \numprint{19.5622} \\
\cellcolor{lightergray}\texttt{RPath}$_{d=20, r=3}$ & \numprint{0.373532}  & \numprint{21.2306} & \\
                  \bottomrule

\end{tabular}
\label{tah:paretoconfigurations}
\end{table*}

\paragraph*{Breadth First Search.} The breadth first search-based algorithm has only one parameter, which is the maximum allowed depth of the breadth first search that searches for a lower degree vertex. We varied the depth $d$ for $d \in \{1, 2, 3, 10, 20\}$, and name the corresponding algorithm $\texttt{BFS}_{d}$. The parameter is such that if $d$ is increased, then quality improves, but also running time is increased. A performance profile regarding quality can be found in Fig.~\ref{fig:perfbfs}. As expected, it can be observed that larger search spaces yield better solutions. While initially this improves quality significantly, i.e.~when increasing the depth 1 to 2, the effect gets less pronounced for larger values. 
For example, on average solution quality improves by 59\% when using \texttt{BFS}$_2$ instead of \texttt{BFS}$_1$, while it only improves by an additional 6.7\% from using \texttt{BFS}$_3$ instead of $\texttt{BFS}_2$. 
The most expensive algorithm in terms of running time in this experiment, $\texttt{BFS}_{20}$, improves solution quality by 2\% over $\texttt{BFS}_{10}$, by 6\% over $\texttt{BFS}_3$, by 13\% over \texttt{BFS}$_2$ and by 80\% over $\texttt{BFS}_1$.  
As the performance profile shows, $\texttt{BFS}_1$ matches the best solution in 2\% of the cases, $\texttt{BFS}_2$ in 18\% of the cases, $\texttt{BFS}_3$ in 55\% of the cases, $\texttt{BFS}_{10}$  in 91\% of the cases, and $\texttt{BFS}_{20}$ always computes the best result. In terms of running time, $\texttt{BFS}_1$ is as expected always the fastest algorithm, being a factor 2.4, 3.8, 10.5, 18.6, faster than $\texttt{BFS}_2$, $\texttt{BFS}_3$, $\texttt{BFS}_{10}$, $\texttt{BFS}_{20}$, respectively. It is also noteworthy that $\texttt{BFS}_3$ computes results on 90\% of the instances that are only 17\% worse than the best solution. On the other hand, it is a factor 5 faster than $\texttt{BFS}_{20}$.

\paragraph*{Random Path.} The random path-based algorithm has two parameters, which are the maximum allowed depth of the random path search, as well as the number of repetitions if the random path search was not successful. In this section, we varied the depth of the algorithm by $d\in\{1,2,3,10,20,50\}$ and the number of repetitions per insertion $r \in \{1,2,3,4,10\}$. In this case, both parameters are exhaustive, in the sense that larger parameters yield better results, but also increase running time.  As this space already spans 30 different algorithm configurations, we computed the Pareto configurations in terms of average quality and running time. This, however, still gives us 24 different configurations on the Pareto front. We interpret this as each configuration gives a different trade-off for quality and running time while none of the parameters $d$ and $r$ is more important. Thus we pick representatives of the more expensive, as well as faster algorithms and algorithms in the middle  from the set of Pareto configurations. 
We use $\texttt{RPath}_{d=50, r=10}, \texttt{RPath}_{d=50, r=4}$ as expensive algorithms, $\texttt{RPath}_{d=50, r=2}$ and $\texttt{RPath}_{d=20, r=3}$  as algorithms in the middle, and the algorithms $\texttt{RPath}_{d=2, r=1}, \texttt{RPath}_{d=1, r=2}$ as fast algorithms. Figure~\ref{fig:exprandompath} shows performance profiles for these algorithms.
The fastest  and lowest quality algorithm is  $\texttt{RPath}_{d=1, r=2}$. 
It is a factor 1.1, 4.4, 4.6, 7.1, 12.5, faster on average than  $\texttt{RPath}_{d=2, r=1}$, $\texttt{RPath}_{d=20, r=3}$, $\texttt{RPath}_{d=50, r=2}$, $\texttt{RPath}_{d=50, r=4}$, $\texttt{RPath}_{d=50, r=10}$, respectively. 
The slowest and highest quality algorithm is $\texttt{RPath}_{d=50, r=10}$. It computes 2.2\%, 4.9\%, 4.1\%, 63.1\% and 68.3\% better solutions than $\texttt{RPath}_{d=50, r=4}$, $\texttt{RPath}_{d=50, r=2}$, $\texttt{RPath}_{d=20, r=3}$, $\texttt{RPath}_{d=2, r=1}$, $\texttt{RPath}_{d=1, r=2}$, respectively.

\begin{figure*}[t!]
\centering
\includegraphics[width=8.25cm]{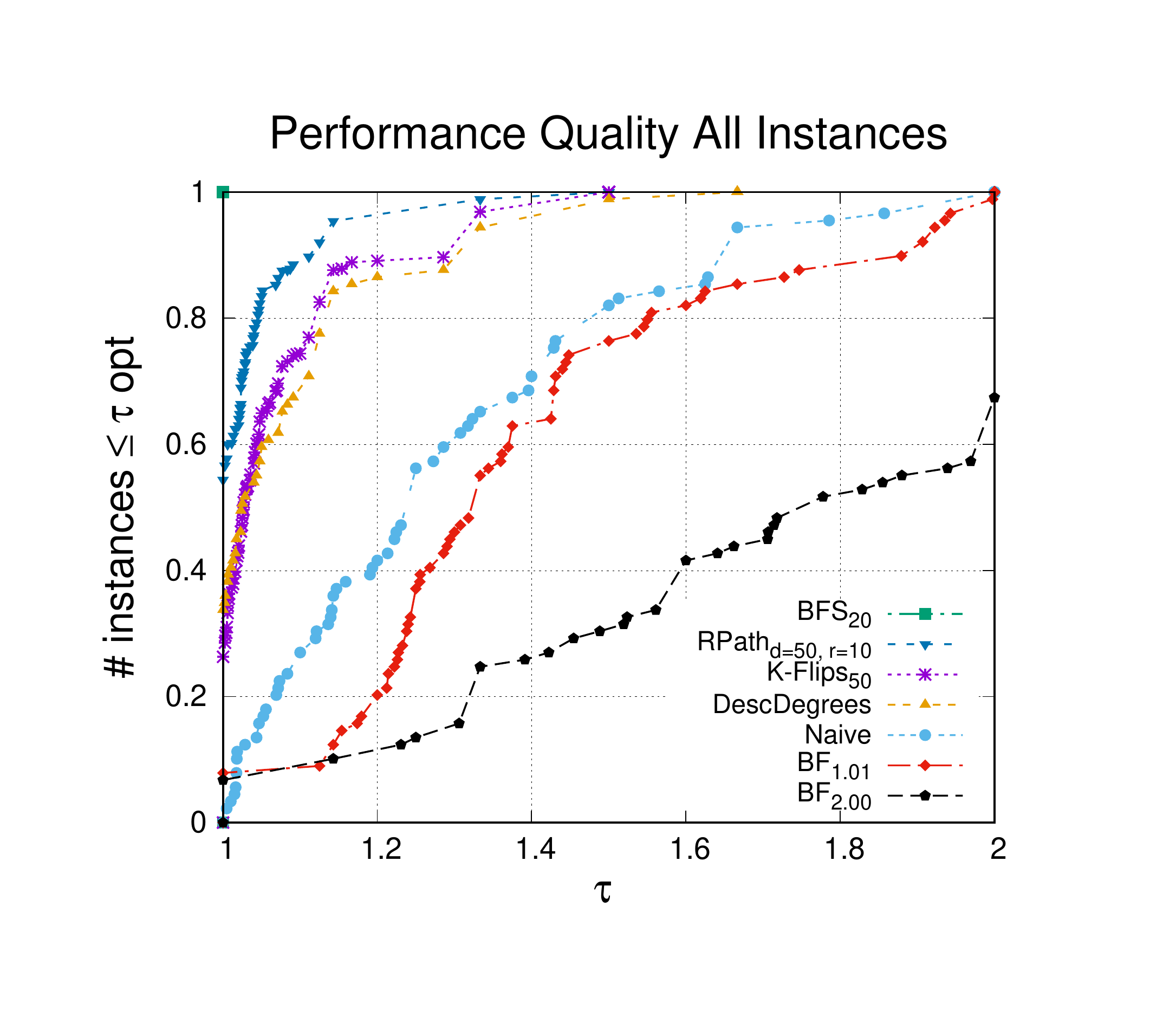}
\includegraphics[width=8.25cm]{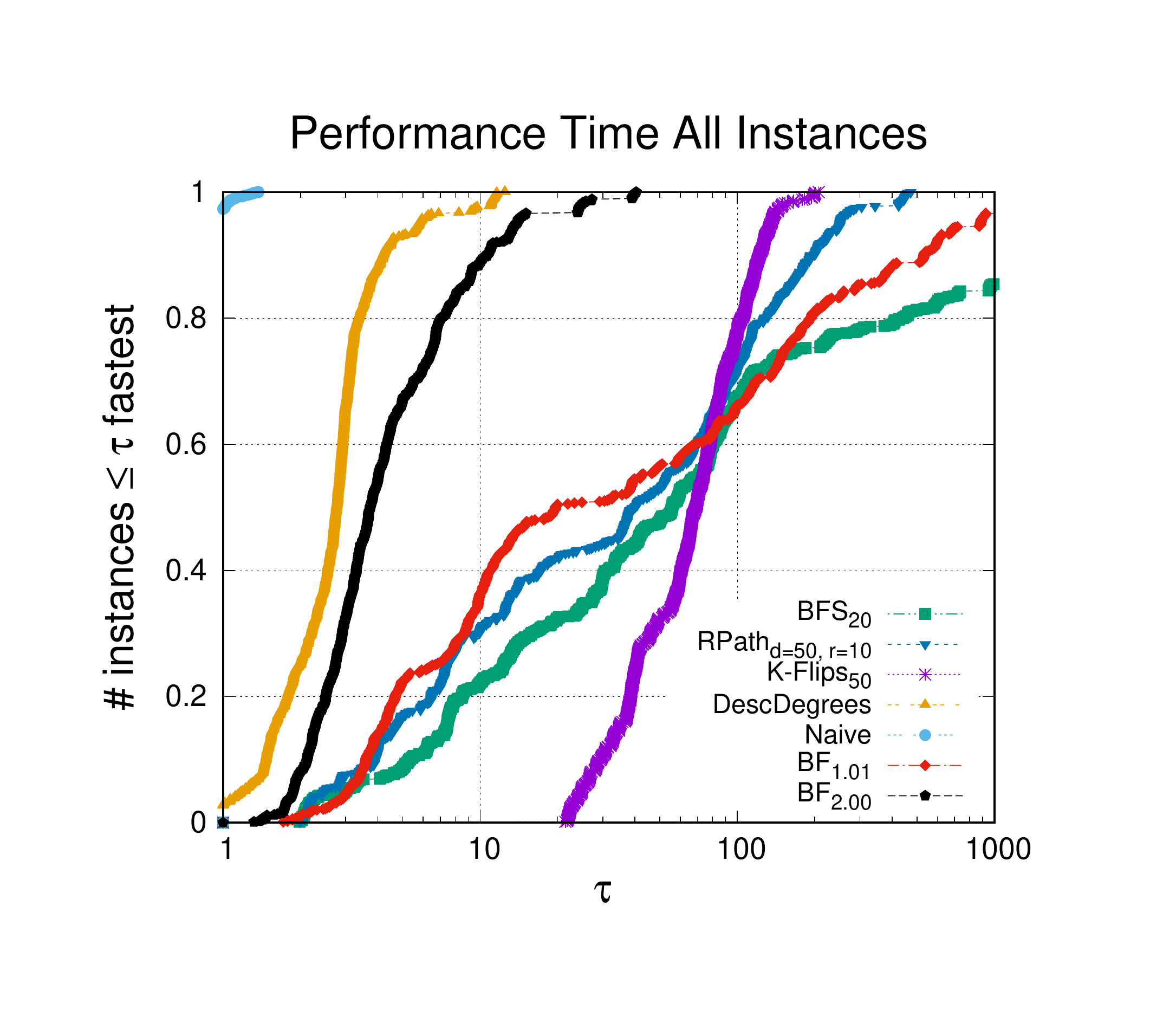} \\
                       \vspace*{-1cm}

\includegraphics[width=8.25cm]{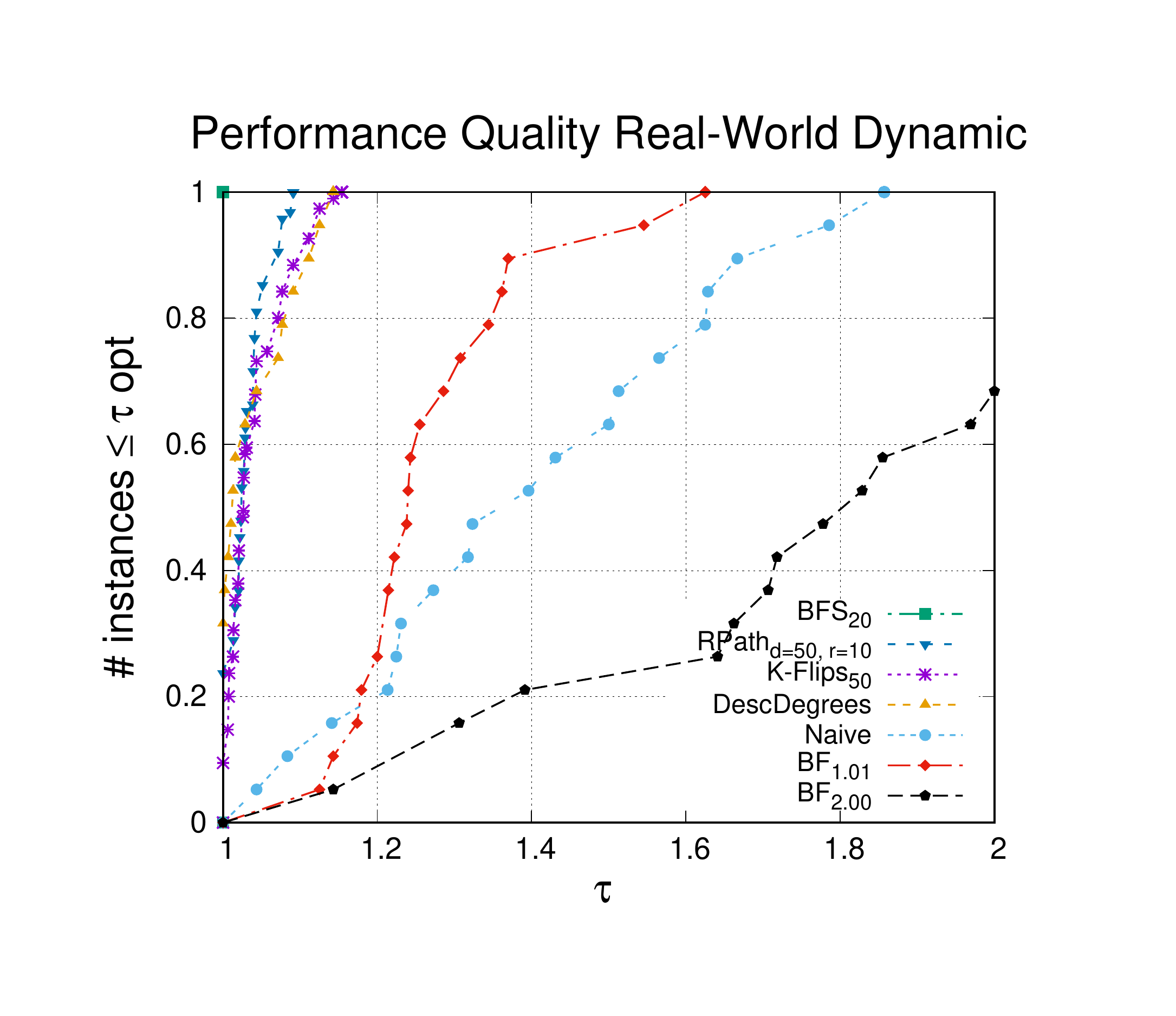}
\includegraphics[width=8.25cm]{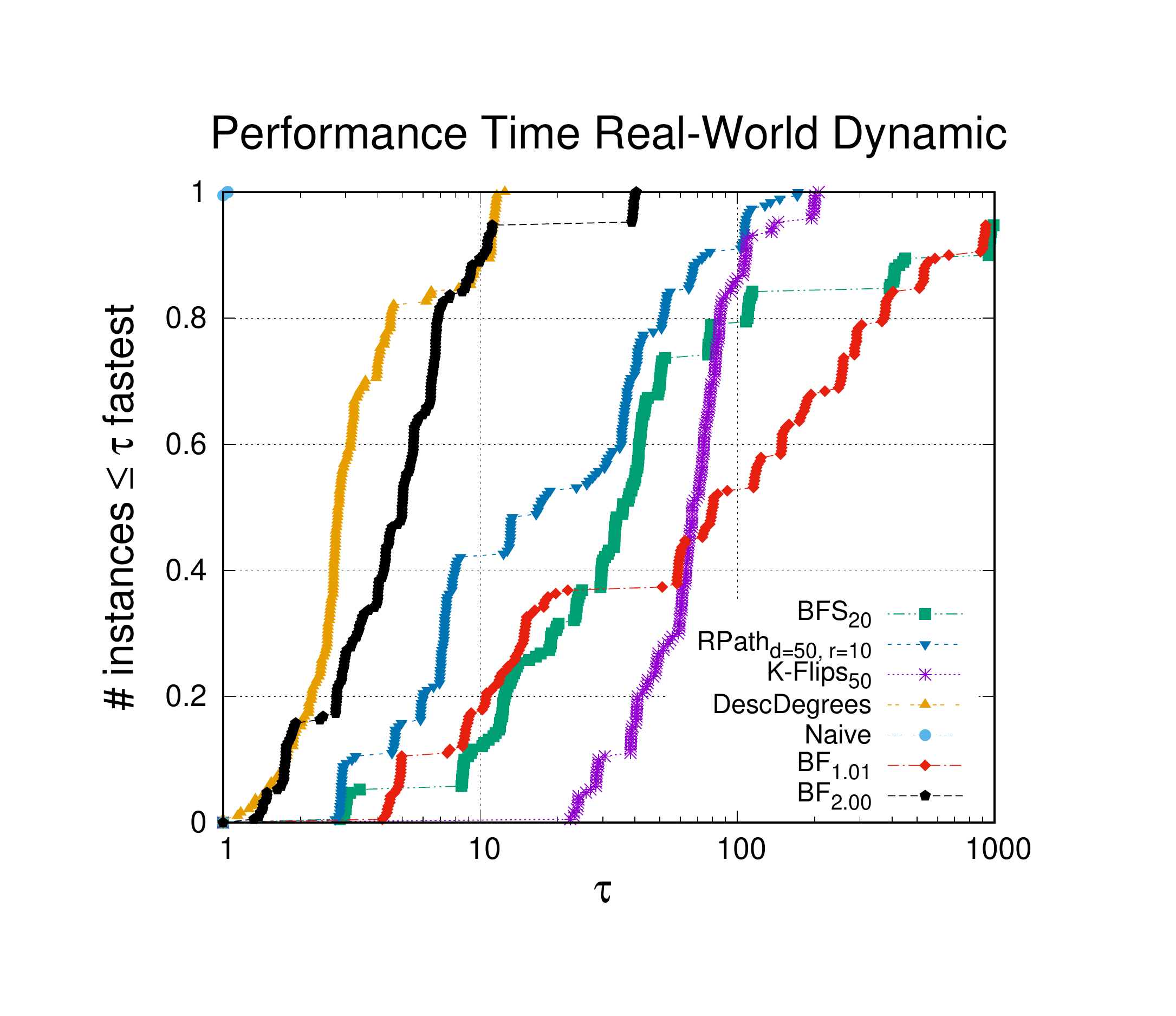} \\
                       \vspace*{-1cm}

\caption{Performance profiles for selected algorithms for quality (left) and time (right), as well as on all instances (top) and only real-world dynamic instances (bottom).}
\label{fig:perfallxyz}
\end{figure*}

\paragraph*{$K$-Flips.} The $K$-Flips algorithm has one parameter~$k$ which is the number of flips that are performed per update. We use $k \in \{1,2,3,10,20,50\}$ as number of flips per update in this experiment, and name the corresponding algorithm $\texttt{K-Flips}_k$. Figure~\ref{fig:expkflips} gives performance profiles for the different algorithm configurations. As for \texttt{BFS}, the parameter of the K-Flips algorithm is exhaustive, i.e.~a larger parameter typically yields better solutions at the expense of additional running time. For example, $\texttt{K-Flips}_{50}$ on average computes 0.6\%, 1.3\%, 3.5\%, 5.0\% and 10.1\%  better solutions than $\texttt{K-Flips}_{20}$, $\texttt{K-Flips}_{10}$, $\texttt{K-Flips}_{3}$, $\texttt{K-Flips}_{2}$, and $\texttt{K-Flips}_{1}$, respectively. On the other hand, $\texttt{K-Flips}_{1}$ is a factor 
1.5, 1.9, 4.9, 8.7, 19.8, faster than \texttt{K-Flips}$_2$, \texttt{K-Flips}$_3$, \texttt{K-Flips}$_{10}$, \texttt{K-Flips}$_{20}$,~\texttt{K-Flips}$_{50}$.

\paragraph*{Brodal-Fagerberg.} The algorithm by Brodal and Fagerberg naturally has no parameters. However, we added the parameter $\beta$ that controls the current bound for the arboricity of the graph. We use $\beta \in \{1.01, 1.1, 1.5, 2\}$ in this experiment and denote the corresponding algorithm with $\texttt{BF}_\beta$. Note that $\texttt{BF}_2$ corresponds to the proposed version of the algorithm by Brodal and Fagerberg. Figure~\ref{fig:expbf} gives performance profiles for running time and solution quality. For smaller values of $\beta$, the algorithm computes better results, but has larger running times. This is expected, as the algorithm only tries to maintain an $\alpha$-orientation for the current $\alpha$ and to do that it is fairly aggressive in terms of edge reorientations. If $\beta$ is to large, then the algorithm effectively ignores ``good'' solutions after/during the rebuild and the update of $\alpha$. The increase in running time for smaller values of $\beta$ is due to more rebuilds being triggered. In terms of quality, the algorithm $\texttt{BF}_{1.01}$ computes 2.3\%, 15.9\%, 29.3\% better solutions on average than $\texttt{BF}_{1.1}$, $\texttt{BF}_{1.5}$ and $\texttt{BF}_{2}$, respectively. On the other hand, the original version $\texttt{BF}_2$ is a factor 1.4, 3.5, 7.9 faster than $\texttt{BF}_{1.5}$, $\texttt{BF}_{1.1}$, and $\texttt{BF}_{1.01}$, respectively.

\subsection{Overall Comparison.}
\label{exp:overall}
We now start to compare all of the algorithms studied in this paper against each other. To do so, first we took all of the algorithm configurations from Section~\ref{sec:parameterstudy} and computed their average for quality and time. Afterwards we computed the Pareto profile. Overall average results over all instances, as well as Pareto configurations can be found in Table~\ref{tah:paretoconfigurations}. 

The \texttt{Naive} algorithm is, as expected, the fastest algorithm since the update only performs one comparison and then directly assigns the edge. Surprisingly, this simple strategy is overall already fairly good. The average maximum out degree of the \texttt{Naive} algorithm is 28.0\% higher than that of the best algorithm \texttt{BFS}$_{20}$ while being a factor 67.7 faster. The next fastest algorithm is the \texttt{DescDegrees} algorithm which is a factor 25.8 faster than $\texttt{BFS}_{20}$, while computing solutions that are 8.6\% worse than the best algorithm. We believe that the \texttt{DescDegrees} algorithm performs so well because of the choice to swap an edge with the smallest degree vertex in the neighborhood. This makes it easier for later update steps to find vertices that have room for an additional edge. The random path configurations $\texttt{RPath}_{d=20, r=3}$, $\texttt{RPath}_{d=50, r=2}$ are on par in terms of solution quality with \texttt{DescDegrees} (or slightly better), but need factor 4.1 and 4.3 more time, respectively. The faster configurations \texttt{RPath}$_{d=1,r=2}$, and \texttt{RPath}$_{d=2,r=1}$ that are faster than the \texttt{DescDegrees} algorithm do not seem to be competitive in terms of solution quality. Moreover, \texttt{$K$-Flips}$_{1,2,3}$, $\texttt{BFS}_{1,2}$ and $\texttt{BF}_{2, 1.50}$ are also dominated by the \texttt{DescDegrees} algorithm which is faster and computes better solutions. Note that the $\texttt{$K$-Flips}_1$ algorithm is not faster than the \texttt{Naive} algorithm since it has to maintain a bucket data structure to efficiently be able to access the nodes that currently have maximum degree. 

\noindent \textit{Conclusion Time/Quality Trade-Off:} Overall, if \emph{speed} is important in applications, we \emph{conclude} that the \texttt{Naive} and the \texttt{DescDegrees} algorithms are the method of choice.

In terms of quality, the $\texttt{BFS}$ and $\texttt{RPath}$ algorithms compute the best solutions. The best algorithm is $\texttt{BFS}_{20}$, followed by \texttt{BFS}$_{10}$, $\texttt{RPath}_{d=50, a=10}$, and lastly $\texttt{RPath}_{d=50, a=4}$. The last three algorithms in that list compute solutions that are 2.2\%, 3.5\%  and 5.8\% worse than solutions computed by  $\texttt{BFS}_{20}$ while being a factor 1.8, 2.2 and 3.9 faster, respectively. The $\texttt{$K$-Flips}$ algorithms is also able to compute very good solutions. For example, $\texttt{$K$-Flips}_{50}$ computes solutions that are 7.1\% worse than the quality of the best algorithm while the running time is slightly better. On the other hand, the best version of the \texttt{BF} algorithm computes solutions that are 36.5\% worse than that of the best algorithm. 

Figure~\ref{fig:perfallxyz} shows performance profiles for the best algorithms in terms of quality from each category as well as the original version of the \texttt{BF} algorithm, \texttt{BF}$_2$. 
The profiles are computed on all instances as well as on real-world dynamic instances only. 
The classification of the algorithms looks similar to the one obtained by looking at average data. \texttt{BFS}$_{20}$ overall computes the best results on all instances. The $\texttt{RPath}_{d=50,10}$ algorithm overall computes the best results on 54\% of all the instances, but only computes the best result on 24\% of the real-world dynamic instances. \texttt{DescDegrees} computes the optimum result in 33.7\% of the cases, and in 31.6\% on real-world dynamic instances. The \texttt{$K$-Flips}$_{50}$ algorithms obtains the best result on 9.5\% of all instances, and on 26.2\% of the real-world dynamic instances. The algorithms $\texttt{Naive}$, $\texttt{BF}_{1.01}$, and $\texttt{BF}_2$ compute the best solution in less than 7\% of the cases each.
While $\texttt{BF}_{1.01}$ is worse than the $\texttt{Naive}$ algorithm on all instances, it is better on real-world dynamic instances only. $\texttt{BF}_2$ is not competitive in terms of quality.

\noindent \textit{Conclusion Quality:} We \emph{conclude}, if quality is important in applications, then the \texttt{BFS} algorithms are a good choice with an easy parameter to control quality.
\begin{table}
\centering
\vspace*{-.25cm}
\caption{The percentage of solutions that the respective algorithm did compute an optimum result after all updates have been performed. The percentage is computed only on the instances that the optimum solver could solve.}
\label{tab:optresults}
\vspace*{.25cm}
\begin{tabular}{lr}
\toprule
Algorithm & \# opt solutions\\
                  \midrule
\texttt{BFS}$_{20}$ &  90.6\%                  \\
\texttt{RPath}$_{d=50, r=10}$ & 50.6\%                   \\
\texttt{DescDegrees} & 20.3\%                   \\
\texttt{$K$-Flips}$_{50}$ &  19.1\%                  \\
\texttt{BF}$_{1.01}$ & 4.7\%                  \\
\texttt{Naive} & 0                  \\
        \bottomrule
\end{tabular}
\vspace*{-.65cm}
\end{table}

\paragraph*{Comparison to Optimum Results.} We ran the ILP from Section~\ref{sec:ilp} using Gurobi 9.12 on the instance that is obtained after all insertions and deletion operations have been done. We gave Gurobi a time limit of 10 hours. However, many instances finish within a couple of minutes. Overall, the solver has computed 64 optimum results (out of 83 instances overall) and crashed on the remaining instances due insufficient memory of our machine which has 100GB of RAM. Table~\ref{tab:optresults} shows how many instance the best algorithm configurations from each category could solve to optimality on the instances that the ILP could solve. The $\texttt{BFS}_{20}$ algorithm performs very well -- it computes the optimum result after all updates in 90.6\% of the cases. On average, the algorithm computes solutions that are $2.4$\% worse than the optimum solution. The  \texttt{RPath}$_{d=50, r=10}$  algorithm, which is the best algorithm from the random path algorithms in terms of quality, still computes the optimum result in 50.6\% of the cases. Notably is also the fast \texttt{DescDegrees} algorithm which is able to compute the optimum result in $20.3\%$ of the cases. Roughly, the same result (19.1\%) is achieved by the \texttt{$K$-Flips}$_{50}$ algorithm. The \texttt{BF}$_{1.01}$ algorithm very rarely computes optimum results, the \texttt{Naive} algorithm never computes an optimum result.

\paragraph*{Advice for Practitioners:} 

We summarize: if \emph{speed} is important in applications, then the \texttt{Naive} or the \texttt{DescDegrees} algorithms should be used as they are either the fastest algorithm or a good trade-off of quality and speed. If \emph{quality} is important in applications, then the \texttt{BFS} algorithms should be employed.
Lastly, the experiments conducted in this section show that the new algorithms outperform/dominate the algorithms currently proposed in the literature in terms of quality \emph{and} running time. However, one can also see that the $\texttt{K-Flips}$ algorithms are able to compute very high quality solutions in a reasonable amount of time. Moreover,  the newly proposed currently do not have a guarantee on solution quality. Hence, if in applications a \emph{guarantee} on solution quality is required, we recommend using the $\texttt{$K$-Flips}$ algorithm.

\section{Conclusion}
\ifDoubleBlind
\else
We presented a range of new algorithms as well as implementations of algorithms from the literature for the fully dynamic $\Delta$-orientation problem, which is to maintain an orientation of the edges of the graph such that the out-degree is low. 
While there has been theoretical work on dynamic versions of this problem, there has been no experimental evaluation available. 
We started to close this gap and performed a range of experiments with newly proposed algorithms as well as algorithms from the current literature.
Experiments indicate that the newly developed algorithms dominate algorithms from the current literature.
The best algorithm considered in this paper in terms of quality, $\texttt{BFS}_{20}$ computes the optimum result on over 90\%  of the instances that our optimum solver could solve.
It is on average only 2.4\% worse than the optimum solution.
Given the good results, we plan to release the codes as open source soon. 
In future work, it will be interesting to look at applications of these type of dynamic algorithms such as dynamic edge coloring.
Moreover, it would be interesting to see if the descending degrees as well as the breadth-first search-based algorithms do provide theoretical guarantees.
\fi

\ifDoubleBlind
\else
\section*{Acknowledgements} 
We acknowledge support by DFG grant SCHU 2567/3-1. Moreover, we like to acknowledge Dagstuhl Seminar 22461 on dynamic graph algorithms. %
\fi

\nprounddigits{0}
{
        \footnotesize
\bibliographystyle{plainnat}
}
\bibliography{references}

\begin{appendix}
\section{Instances}
\begin{table*}[h!]
      \centering
      \vspace*{-1cm}
      \caption{Basic properties of benchmark set of static and real dynamic graphs graphs from~\cite{benchmarksfornetworksanalysis,DBLP:journals/corr/abs-2003-00736,UFsparsematrixcollection,snap,DBLP:conf/www/Kunegis13,konect:unlink,DBLP:journals/jpdc/FunkeLMPSSSL19,kappa}.
In case of real-world dynamic graphs we report the original number update operations $\mathcal{O}$ before removing obsolete updates (such as parallel edges, self-loops etc.). Most of these instances only feature insertions. The two exceptions are marked with a *. } 

      \begin{tabular}{lrr@{\hskip 13pt} lrr@{\hskip 13pt} }
              \midrule

\multicolumn{6}{c}{Mesh Type Networks} \\
              \midrule
      graph & $n$ & $m$ & graph & $n$ & $m$ \\
                \midrule
144 &\numprint{144649} & \numprint{1074393} &                fe\_4elt2 &\numprint{11143} & \numprint{32818}  \\
3elt &\numprint{4720} & \numprint{13722} &                     fe\_body &\numprint{45087} & \numprint{163734}  \\
4elt &\numprint{15606} & \numprint{45878} &                    fe\_ocean &\numprint{143437} & \numprint{409593}  \\
598a &\numprint{110971} & \numprint{741934} &                  fe\_pwt &\numprint{36519} & \numprint{144794} \\
add20 &\numprint{2395} & \numprint{7462} &                     fe\_rotor &\numprint{99617} & \numprint{662431} \\ 
add32 &\numprint{4960} & \numprint{9462} &                     fe\_sphere &\numprint{16386} & \numprint{49152}  \\
auto &\numprint{448695} & \numprint{3314611} &                 fe\_tooth &\numprint{78136} & \numprint{452591}  \\
bcsstk29 &\numprint{13992} & \numprint{302748} &               finan512 &\numprint{74752} & \numprint{261120} \\
bcsstk30 &\numprint{28924} & \numprint{1007284} &              m14b &\numprint{214765} & \numprint{1679018} \\
bcsstk31 &\numprint{35588} & \numprint{572914} &               memplus &\numprint{17758} & \numprint{54196} \\
bcsstk32 &\numprint{44609} & \numprint{985046} &               rgg15 &$2^{15}$ & \numprint{160240}\\
bcsstk33 &\numprint{8738} & \numprint{291583} &                rgg16&$2^{16}$& \numprint{342127}\\
brack2 &\numprint{62631} & \numprint{366559} &                 rgg20&$2^{20}$& \numprint{728753}\\
crack &\numprint{10240} & \numprint{30380} &                   t60k &\numprint{60005} & \numprint{89440} \\
cs4 &\numprint{22499} & \numprint{43858} &                     uk &\numprint{4824} & \numprint{6837} \\
cti &\numprint{16840} & \numprint{48232} &                     vibrobox &\numprint{12328} & \numprint{165250}\\
data &\numprint{2851} & \numprint{15093} &                     wave &\numprint{156317} & \numprint{1059331} \\
delaunay16 & $2^{16}$ & \numprint{196575}  &                   whitaker3 &\numprint{9800} & \numprint{28989}  \\
delaunay17 & $2^{17}$ & \numprint{393176} &                    wing &\numprint{62032} & \numprint{121544} \\
delaunay{20} & $2^{20}$ & \numprint{3145686} &                 wing\_nodal &\numprint{10937} & \numprint{75488} \\

                \midrule
\multicolumn{6}{c}{Social Networks} \\
                \midrule
      graph & $n$ & $m$ & graph & $n$ & $m$ \\
                \midrule
amazon-2008 &\numprint{735323} & \numprint{3523472} &         in-2004 &\numprint{1382908} & \numprint{13591473}                      \\
as-22july06 &\numprint{22963} & \numprint{48436} &            loc-brightkite\_edges &\numprint{56739} & \numprint{212945} \\
as-skitter &\numprint{554930} & \numprint{5797663} &          loc-gowalla\_edges &\numprint{196591} & \numprint{950327} \\
citationCiteseer &\numprint{268495} & \numprint{1156647} &    p2p-Gnutella04 &\numprint{6405} & \numprint{29215} \\
cnr-2000 &\numprint{325557} & \numprint{2738969} &            PGPgiantcompo &\numprint{10680} & \numprint{24316}   \\                         
coAuthorsCiteseer &\numprint{227320} & \numprint{814134} &    rhg1G                                      & 100.0M & $\approx$1G  \\
coAuthorsDBLP &\numprint{299067} & \numprint{977676} &        rhg2G                                      & 100.0M & $\approx$2G  \\
coPapersCiteseer &\numprint{434102} & \numprint{16036720}      & soc-Slashdot0902 &\numprint{28550} & \numprint{379445} \\
coPapersDBLP &\numprint{540486} & \numprint{15245729} &                 wordassociation-2011 &\numprint{10617} & \numprint{63788} \\
email-EuAll &\numprint{16805} & \numprint{60260} &            web-Google &\numprint{356648} & \numprint{2093324} \\
enron &\numprint{69244} & \numprint{254449} &                         wiki-Talk &\numprint{232314} & \numprint{1458806} \\

                \midrule
\multicolumn{6}{c}{Real-World Dynamic Networks} \\
                \midrule
      graph & $n$ & $\mathcal{O}$ & graph & $n$ & $\mathcal{O}$ \\
                \midrule
amazon-ratings &\numprint{2146058} & \numprint{5838041}       & movielens10m &\numprint{69879} & \numprint{10000054}                                   \\ 
citeulike\_ui &\numprint{731770} & \numprint{2411819}         & munmun\_digg &\numprint{30399} & \numprint{87627}                                     \\ 
dewiki$^*$ &\numprint{2166670} & \numprint{86337879}       &    proper\_loans &\numprint{89270} & \numprint{3394979}                                  \\ 
dewiki-2013 &\numprint{1532354} & \numprint{33093029}      &    sociopatterns-infections &\numprint{411} & \numprint{17298}                            \\ 
dnc-temporalGraph &\numprint{2030} & \numprint{39264}      &    stackexchange-stackoverflow &\numprint{545197} & \numprint{1301942}                    \\ 
facebook-wosn-wall &\numprint{46953} & \numprint{876993}   &    topology &\numprint{34762} & \numprint{171403}                                                 \\ 
flickr-growth &\numprint{2302926} & \numprint{33140017}      &  wikipedia-growth &\numprint{1870710} & \numprint{39953145}                            \\ 
haggle &\numprint{275} & \numprint{28244}                    &  wiki\_simple\_en$^*$ &\numprint{100313} & \numprint{1627472}                          \\ 
lastfm\_band &\numprint{174078} & \numprint{19150868}        &  youtube-u-growth &\numprint{3223590} & \numprint{9375374}                              \\ 
lkml-reply &\numprint{63400} & \numprint{1096440}            &                    \\ 

      \bottomrule
      \end{tabular}
      \label{staticgraphs}
\label{dyninstances}
\label{tab:graphstable}
\end{table*}

\end{appendix}

\end{document}